\documentclass{aa}

\usepackage{graphicx}
\usepackage{txfonts}
\usepackage{bbm}
\usepackage{url}
\usepackage{natbib}
\usepackage[breaklinks=true]{hyperref} 
\hypersetup{colorlinks=true,urlcolor=blue,citecolor=blue,pdfborder= 0 0 0}
\bibpunct{(}{)}{;}{a}{}{,}    
\begin{document}

   \title{Probabilistic fibre-to-target assignment algorithm for multi-object spectroscopic surveys}


   \author{E.~Tempel\inst{1} 
          \and
                  P.~Norberg\inst{2} 
                  \and
                  T.~Tuvikene\inst{1} 
                  \and
                  T.~Bensby\inst{3} 
                  \and
                  C.~Chiappini\inst{4} 
                  \and
                  N.~Christlieb\inst{5} 
                  \and
                  M.-R.~L.~Cioni\inst{4} 
                  \and
                  J.~Comparat\inst{6} 
                  \and
                  L.~J.~M.~Davies\inst{7} 
                  \and
                  G.~Guiglion\inst{4} 
                  \and
                  A.~Koch\inst{8} 
                  \and
                  G.~Kordopatis\inst{9} 
                  \and
                  M.~Krumpe\inst{4} 
                  \and
                  J.~Loveday\inst{10} 
                  \and
                  A.~Merloni\inst{6} 
                  \and
                  G.~Micheva\inst{4} 
                  \and
                  I.~Minchev\inst{4} 
                  \and
                  B.~F.~Roukema\inst{11,12} 
                  \and
                  J.~G.~Sorce\inst{4,12} 
                  \and
                  E.~Starkenburg\inst{4} 
                  \and
                  J.~Storm\inst{4} 
                  \and
                  E.~Swann\inst{13} 
                  \and
                  W.~F.~Thi\inst{6} 
                  \and
                  G.~Traven\inst{3} 
                  \and
                  R.~S.~de~Jong\inst{4} 
          }

   \institute{
   Tartu Observatory, University of Tartu, Observatooriumi 1, 61602 T\~oravere, Estonia\\
              \email{elmo.tempel@ut.ee}
                          \and
                          Institute for Computational Cosmology and Centre for Extragalactic Astronomy, Department of Physics, Durham University, South Road, Durham DH1 3LE, UK
                          \and
                          Lund Observatory, Department of Astronomy and Theoretical Physics, Box~43, SE-221~00 Lund, Sweden
                          \and
                          Leibniz-Institut f\"ur Astrophysik Potsdam (AIP), An der Sternwarte 16, D-14482 Potsdam, Germany
                          \and
                          Zentrum f\"ur Astronomie der Universit\"at Heidelberg, Landessternwarte, K\"onigstuhl 12, 69117 Heidelberg, Germany
                          \and
                          Max-Planck-Institut f\"ur Extraterrestrische Physik (MPE), Giessenbachstraße, D-85748 Garching, Germany
                          \and
                          ICRAR, The University of Western Australia, 35 Stirling Highway, Crawley, WA 6009, Australia
                          \and
                          Zentrum f\"ur Astronomie der Universit\"at Heidelberg, Astronomisches Rechen-Institut, M\"onchhofstr. 12, 69120 Heidelberg, Germany
                          \and
                          Université Côte d’Azur, Observatoire de la Côte d’Azur, CNRS, Laboratoire Lagrange, France
                          \and
                          Astronomy Centre, University of Sussex, Falmer, Brighton BN1 9QH, UK
                          \and
                          Institute of Astronomy, Faculty of Physics, Astronomy and Informatics, Nicolaus Copernicus University, Grudziadzka 5, 87-100 Toru\'n, Poland
                          \and
                          Université Lyon 1, Ens de Lyon, CNRS, Centre de Recherche Astrophysique de Lyon UMR5574, 69230 Saint-Genis-Laval, France
                          \and
                          Institute of Cosmology and Gravitation, University of Portsmouth, Burnaby Road, Portsmouth PO1 3FX, UK
             }

   \date{Received 02 December 2019 / Accepted 28 January 2020}

  \abstract
   {Several new multi-object spectrographs are currently planned or under construction that are capable of observing thousands of Galactic and extragalactic objects simultaneously.}
   {In this paper we present a probabilistic fibre-to-target assignment algorithm that takes spectrograph targeting constraints into account and is capable of dealing with multiple concurrent surveys. We present this algorithm using the 4-metre Multi-Object Spectroscopic Telescope (4MOST) as an example.}
   {The key idea of the proposed algorithm is to assign probabilities to fibre-target pairs. The assignment of probabilities takes the fibre positioner's capabilities and constraints into account. Additionally, these probabilities include requirements from surveys and take the required exposure time, number density variation, and angular clustering of targets across each survey into account. The main advantage of a probabilistic approach is that it allows for accurate and easy computation of the target selection function for the different surveys, which involves determining the probability of observing a target, given an input catalogue.}
   {The probabilistic fibre-to-target assignment allows us to achieve maximally uniform completeness within a single field of view. The proposed algorithm maximises the fraction of successfully observed targets whilst minimising the selection bias as a function of exposure time. In the case of several concurrent surveys, the algorithm maximally satisfies the scientific requirements of each survey and no specific survey is penalised or prioritised.}
   {The algorithm presented is a proposed solution for the 4MOST project that allows for an unbiased targeting of many simultaneous surveys. With some modifications, the algorithm may also be applied to other multi-object spectroscopic surveys.}


   \maketitle
%

\section{Introduction}

Currently, there are many new multi-object spectroscopic facilities planned or undergoing construction. These include the forthcoming wide-field multi-object spectrographs for the William Herschel Telescope \citep[WEAVE,][]{2012SPIE.8446E..0PD}, the Dark Energy Spectroscopic Instrument \citep[DESI,][]{2016arXiv161100036D}, the Subaru Prime Focus Spectrograph \citep[PFS,][]{2016SPIE.9908E..1MT}, the Maunakea Spectroscopic Explorer \citep[MSE,][]{2019arXiv190404907T}, the Multi-Object Optical and Near-infrared Spectrograph \citep[MOONS,][]{2016ASPC..507..109C}, and the VISTA 4-metre Multi-Object Spectroscopic Telescope \citep[4MOST,][]{2019Msngr.175....3D}. All of these instruments are capable of observing thousands of objects simultaneously, allowing the community to carry out several surveys at the same time. To maximise instrument efficiency, several Galactic and extragalactic science cases are addressed in tandem; during a single science exposure, targets from several different surveys are observed simultaneously. This parallel mode of operations poses new challenges, such as how to prioritise different surveys, which were not encountered by previously undertaken surveys that usually only had one main science goal.

In this paper we address this challenge and propose a probabilistic fibre-to-target assignment algorithm for multi-object spectroscopic surveys. A related challenge is the generation of an optimal tiling pattern that will be treated in a separate work (Tempel et al. in prep). The proposed targeting algorithm is currently being developed for the 4MOST survey, but it can be generalised to all multi-object spectroscopic surveys. The algorithm is one potential approach for 4MOST, and its development is an important step forward in defining the best strategy for 4MOST observations. However, at this stage the algorithm is a proposed one and ultimately the implemented algorithm for 4MOST may slightly differ from the one described in this paper.

The 4MOST is a new high-multiplex, wide-field spectroscopic survey facility under development for the four-metre-class Visible and Infrared Survey Telescope for Astronomy (VISTA) at Paranal. 4MOST has a large field of view of 4.3 square degrees and a high multiplex capability, with 1624 fibres feeding two low-resolution spectrographs and 812 fibres transferring light to the high-resolution spectrograph. Spectrograph specifications are described in \citet{2015AN....336..665H}. An overview of the 4MOST project is given by \citet{2019Msngr.175....3D}, scientific operations are described in \citet{2019Msngr.175...12W}, and the current survey strategy is described in \citet{2019Msngr.175...17G}. The 4MOST consortium survey includes the following ten surveys; the Milky Way Halo Low-Resolution Survey \citep{2019Msngr.175...23H}, the Milky Way Halo High-Resolution Survey \citep{2019Msngr.175...26C}, the Milky Way Disc and Bulge Low-Resolution Survey \citep[4MIDABLE-LR,][]{2019Msngr.175...30C}, the Milky Way Disc and Bulge High-Resolution Survey \citep[4MIDABLE-HR,][]{2019Msngr.175...35B}, the eROSITA Galaxy Cluster Redshift Survey \citep{2019Msngr.175...39F}, the Active Galactic Nuclei Survey \citep{2019Msngr.175...42M}, the Wide-Area VISTA Extragalactic Survey \citep[WAVES,][]{2019Msngr.175...46D}, the Cosmology Redshift Survey \citep[CRS,][]{2019Msngr.175...50R}, the One Thousand and One Magellanic Fields Survey \citep[1001MC,][]{2019Msngr.175...54C}, and the Time-Domain Extragalactic Survey \citep[TiDES,][]{2019Msngr.175...58S}. In addition to these 4MOST consortium surveys, additional public surveys will have to be included into the whole 4MOST project planning. The first five-year survey will start at the end of 2022.
        
One of the challenges for multi-object spectroscopic instruments to conduct several surveys in parallel is the computation of target selection functions. In 4MOST, each fibre's home (resting) position in the focal plane is fixed where each fibre has a limited patrol area in which it can move\footnote{4MOST uses the Echidna fibre-positioning technology \citep{2014SPIE.9151E..1XS,2018SPIE10702E..79B}.}.
At any position in the sky, there can be a large choice of targets that can be observed by a specific fibre. The actual strategy to choose the targets has to fulfil certain requirements. Thus, the selection function at small angular scales is sophisticated and this significantly complicates any statistical analysis of observed data. For example, in order to recover the two-point correlation function, assuming that one already knows the sky-plane and line-of-sight selection functions of the target catalogue, requires a detailed knowledge and understanding of the targeting probability for all individual targets and target pairs. 
\citet{2017MNRAS.472.1106B} developed an advanced algorithm to exactly recover the two-point correlation function and they applied it to the DESI dark time surveys \citep{2018MNRAS.481.2338B}, in particular, with \citet{2019MNRAS.484.1285S} considering its use within the DESI Bright Galaxy Survey. The probabilistic fibre-to-target assignment algorithm proposed in this paper allows us to use a similar approach in a straightforward manner. 

The algorithm proposed in the current paper was specifically developed to meet the following conditions. First, the proposed algorithm should be able to recover the targeting selection function. Second, the selection of targets from the input catalogues should be random and the algorithm should achieve a nearly uniform completeness as a function of the required exposure time of targets. Third, the algorithm should balance the surveys observed in parallel while taking the completeness goals of each individual survey into account. The algorithm we propose in this current paper covers all three of these aspects.

The structure of the paper is as follows. In Sect.~\ref{sec:target_allocation} we describe the probabilistic fibre-to-target assignment algorithm in detail. We test the algorithm by using Poisson distributed targets in Sect.~\ref{sec:poisson_test} and then by using mock survey catalogue targets in Sect.~\ref{sec:example_test}. We conclude the paper and discuss further improvements in Sect.~\ref{sec:conclusion}.

\begin{figure*}
\centering
\includegraphics[width=180mm]{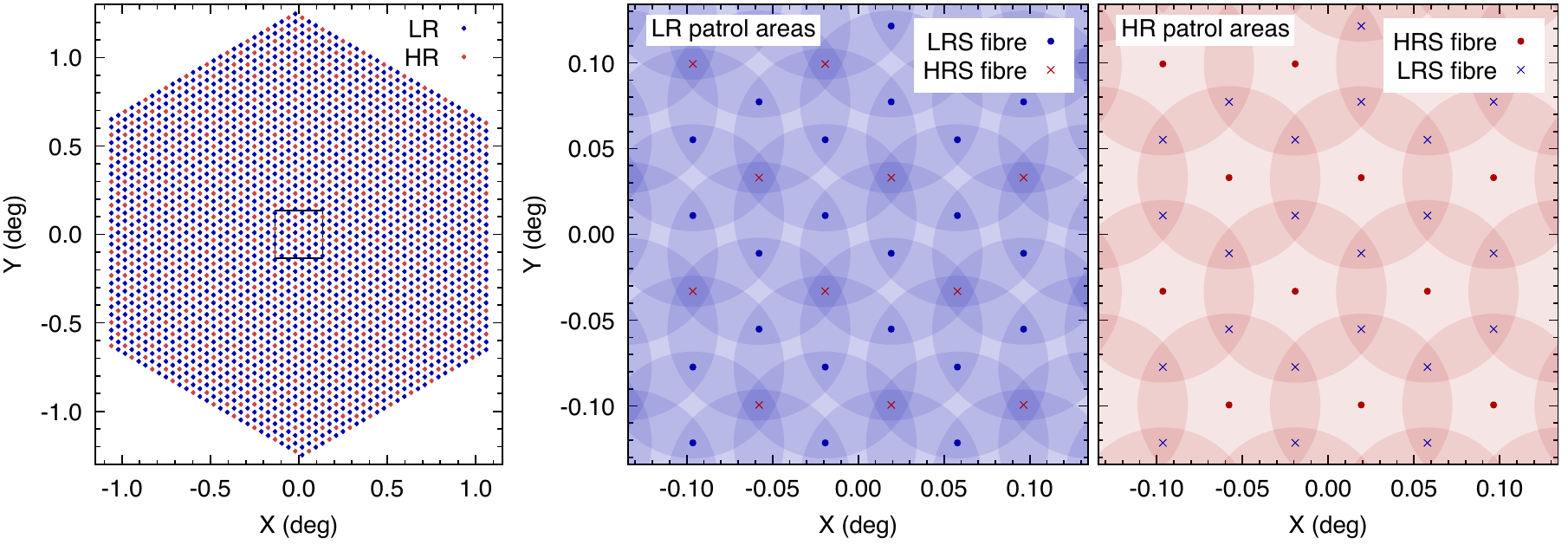}
        \caption{\emph{Left panel}: 4MOST field of view with 1624 low-resolution (blue) and 812 high-resolution spectrograph fibres (red) at their home positions. One field of view covers ~4.3 square degrees in the sky (see also Fig.~\ref{fig:fov_completeness}). Black rectangle depicts the area shown in the middle and right panels. \emph{Middle} and \emph{right} panels show the patrol area (3.2 arcmin from the fibre home position) of each low- and high-resolution spectrograph fibre, respectively. Fibre home positions are marked as points for a given resolution or as small crosses for fibres with alternative resolution. Any location in a field of view can be reached by two to six low-resolution spectrograph fibres and by one to three high-resolution spectrograph fibres. Table~\ref{table:fib_coverage} gives the fraction of area that is covered by a given number of fibres.}
        \label{fig:fibres_multi}
\end{figure*}

\section{Probabilistic targeting algorithm}
\label{sec:target_allocation}

In this section we present the probabilistic fibre-to-target assignment algorithm, using the 4MOST project with its numerous surveys as an example. The presentation of the algorithm is organised as follows. In Sect.~\ref{sec:setup} we describe the requirements and assumptions of the algorithm and define the problem. Sect.~\ref{sec:fibtar_pair} defines how we assign an initial probability for each target-fibre pair and in Sect.~\ref{sec:targeting_algorithm} we describe how the previously assigned probabilities are altered to balance the number densities of targets and how to take survey completeness goals into account. At this point, we have assigned a probability for each fibre-target pair that is used to assign targets for each fibre. Finally, Sect.~\ref{sec:fibtotar} gives an overview about the target allocation scheme used by the algorithm described in the current paper.

\subsection{Set-up of the problem}
\label{sec:setup}

    The 4MOST instrument is a multi-fibre spectrograph with the following constraints. Firstly, the field of view is a hexagon covering 4.3 square degrees of the sky in a single pointing. Secondly, fibre positions across the field of view are fixed in a regular pattern. Each fibre can move only within a small patrol area of radius $r\sim3.2$~arcmin (see Fig.~\ref{fig:fibres_multi}). Thirdly, one third of the fibres are feeding a high-resolution spectrograph (HRS), whilst two-thirds are feeding a low-resolution spectrograph (LRS). During each exposure, high- and low-resolution spectrographs fibres are used simultaneously.
        
        The 4MOST project consists of several consortium surveys and their own sub-surveys of which each is characterised by a specific target selection. The sub-surveys share the focal plane and are observed in parallel as if they are one survey. The challenges the targeting algorithm has to address are the following:
        \begin{enumerate}
                \item As the focal plane is shared by several sub-surveys, objects from multiple sub-surveys are targeted during the same exposure. This is necessary as the target density in the majority of sub-surveys is not high enough to efficiently fill up all fibres in one pointing. In some fields, several sub-surveys require less than a few percent of fibres per pointing, whilst other sub-surveys can fill up all the fibres.
                \item Exposure time requests for different targets vary significantly across each sub-survey. However, due to the above mentioned point, targets with differing exposure time requests must be observed simultaneously. Hence, for a large fraction of targets, repeated observations are necessary to reach the requested total exposure times. Depending on their properties, different targets can be observed more or less efficiently in bright, grey and dark sky conditions; for example, some targets can be easily observed during bright time, whilst in the same field other targets require a long exposure during dark time.
                \item For some sub-surveys, a pre-defined fraction of targets is sufficient to fulfil the sub-survey's science goals. Hence, not all targets in the input catalogue need to be observed. This poses a challenge as to how targets can be selected in a way that all sub-surveys are successfully completed and the required fraction of targets observed.
        \end{enumerate}

It is clear that the target allocation problem in 4MOST is much more complicated than in surveys with only one target class (or single survey). In such surveys, the main problem that occurs is how to most effectively observe the given list of targets. In 4MOST, the problem is how to observe the right set of targets so that the science output of each sub-survey is maximised. Since the mix of sub-surveys and the total target density changes across the sky, there is need for a flexible solution. The solution proposed for the 4MOST survey is a probabilistic fibre-to-target assignment scheme. The key idea is that for each target, the probability that the target is selected for observation by a given fibre is assigned. A detailed description of the probability assignment is presented below.

\begin{table}
\caption{Fraction of area in the 4MOST field of view that can be reached only by $N_\mathrm{fib}$ fibres (left side of the table) or at least with $N_\mathrm{fib}$ fibres (right side of the table). We show the number of fibres that feed the low-resolution spectrograph (LRS) and high-resolution spectrograph (HRS) separately. Given fractions ignore the edges of the field of view, where most of the area can be observed only with one fibre. See Fig.~\ref{fig:fibres_multi} for the illustration of the 4MOST field of view.}
\label{table:fib_coverage}
\centering
\begin{tabular}{c c c | c c c}
\hline\hline
$N_\mathrm{fib}$ & LRS fib. & HRS fib. &
$N_\mathrm{fib}$ & LRS fib. & HRS fib.\\
\hline
   &   \%   & \%    & &  \%   & \% \\
\hline
        1  &  --    & 39.8 & $\geq 1$  &  100.0 & 100.0 \\
        2  &   7.2  & 46.1 & $\geq 2$  &  100.0 & 60.2 \\
        3  &  50.9  & 14.1 & $\geq 3$  &  92.8  & 14.1 \\
        4  &  33.0  & --   & $\geq 4$  &  41.9  & 0.0 \\
        5  &   3.6  & --   & $\geq 5$  &   8.9  & 0.0 \\
        6  &   5.3  & --   & $\geq 6$  &   5.3  & 0.0 \\
\hline
\end{tabular}
\end{table}

\begin{figure*}
\centering
\includegraphics[width=180mm]{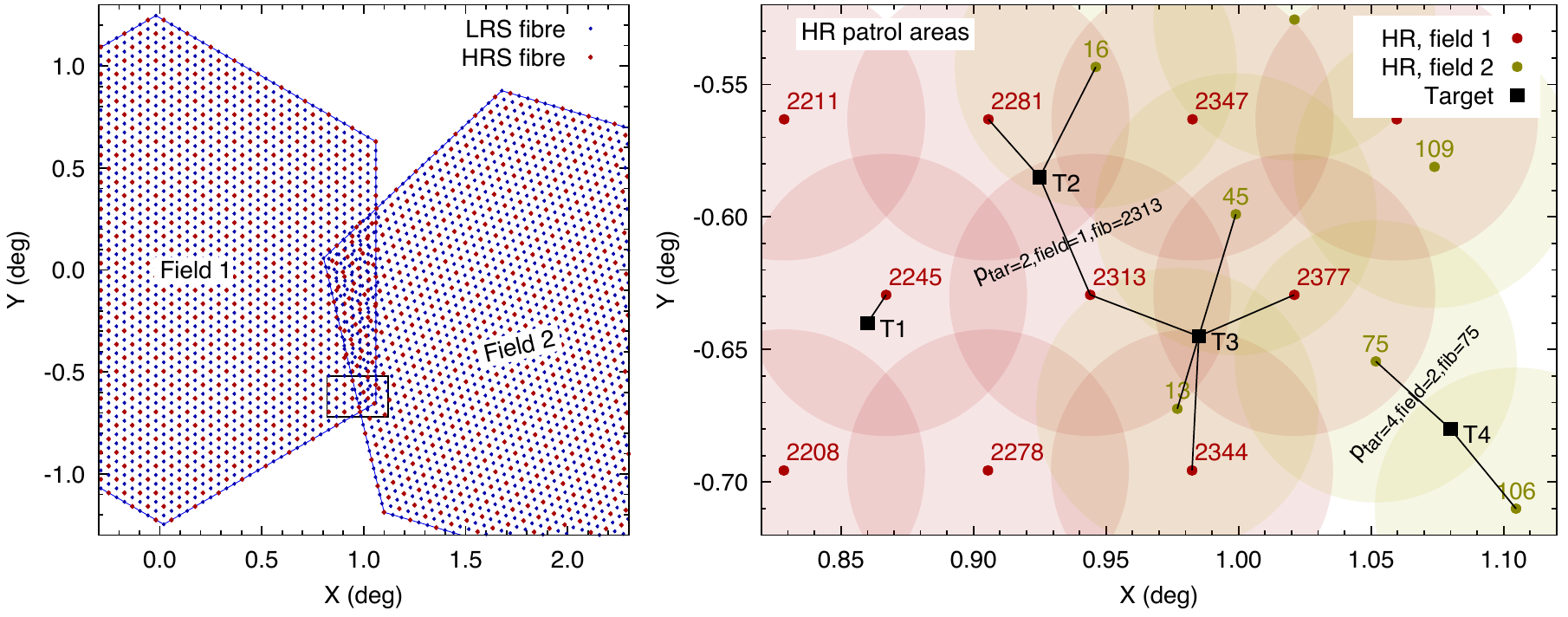}
        \caption{\emph{Left panel}: Two partially overlapping 4MOST fields with low-resolution (blue) and high-resolution (red) spectrograph fibres at their home positions. \emph{Right panel:} Patrol areas of high-resolution spectrograph fibres and four targets in a zoom in region highlighted in left-hand panel. Fibre home positions are marked as points and labelled with fibre numbers. Possible fibre-target pairs are illustrated with lines. Targets T2 and T3 can be reached by fibres from two fields; target T3 can be reached by three fibres from Field 1 and two fibres from Field 2; fibre 2313 can reach two targets (T2 and T3). Sect.~\ref{sec:fibtar_pair} and Eq.~\eqref{eq:ptar_eq1} define the probability $p_{\mathrm{tar},\mathrm{field},\mathrm{fib}}$ for each fibre-target pair.}
        \label{fig:fibre_target}
\end{figure*}

\subsection{Assigning probabilities for fibre-target pairs}
\label{sec:fibtar_pair}

In this section we describe how we assign a probability for a given fibre-target pair. Also, we explain how these probabilities are used and how a target is selected for each fibre are described in Sects.~\ref{sec:targeting_algorithm} and \ref{sec:fibtotar}.

The probabilistic fibre-to-target assignment assumes that for each fibre-target pair, we can assign a probability that a given target should be observed with a given fibre. In order to be able to assign probabilities for each fibre-target pair, the following assumptions have to be met:
\begin{enumerate}
        \item The field pattern (pointing or tile centres as well as orientations) is fixed\footnote{Finding an optimal tiling pattern is an independent algorithm that is described in a separate paper. The tiling algorithm is a critical part of the 4MOST survey optimisation. It essentially decides how many times a given sky region is observed and what is the summed exposure time in a given sky region.}. Before observations begin, we know the centre and orientation of all fields that are to be observed. This is required, as it is necessary to know how many fibres for each target, across all fields that cover the target, could potentially reach this target as well as the total exposure time available for the target.
        \item We know the exposure time and sky condition\footnote{Sky brightness condition is ignored in the current paper and we impose the same sky brightness everywhere, meaning that requested exposure time per target does not depend on the sky brightness.} for each field (for a single science exposure). These are used to compute probabilities for each fibre-target pair. However, during actual observations, the exposure time of a field for a given set of targets can be adjusted with respect to the real observing conditions, such as sky brightness, seeing, and airmass, in order to increase survey efficiency.
        \item For each target we know the required exposure time as a function of the sky condition.
        \item We assume that all fields (generated tiles) in a given sky area are completed by the end of the survey. Since probabilities for fibre-target pairs are assigned using all fields over a given area, the assumption is that all fields are completed. As a result, the proposed algorithm strongly prefers the completion of individual sky areas versus half-completing the entire sky. However, the algorithm does not require that generated fields should allow us to observe all required targets in the input catalogue. In general, the algorithm assumes that the generated field pattern is optimised based on the input target catalogue. 
\end{enumerate}

The probabilistic targeting algorithm starts by performing a probability assignment for each fibre-target pair. For each target $t$ and fibre $k$ in a field $i$, we assign a probability as follows (see Fig.~\ref{fig:fibre_target} for illustration):
\begin{equation}
        p_{\mathrm{tar}=t, \mathrm{field}=i, \mathrm{fib}=k} = c_\mathrm{tar}(t,i) \cdot
         p^{\mathrm{raw}}_{\mathrm{tar}=t, \mathrm{field}=i, \mathrm{fib}=k} ,
         \label{eq:ptar_eq1}
\end{equation}
\begin{align}
        &p^\mathrm{raw}_{\mathrm{tar}=t, \mathrm{field}=i, \mathrm{fib}=k} = 
        \mathbbm{1}\!\left\{ \mathrm{Res}_{\mathrm{tar}=t} = \mathrm{Res}_{\mathrm{fib}=k} \right\} \cdot f_\mathrm{tilt}(t,i,k) 
        \cdot  \nonumber \\ &\qquad\qquad f_\mathrm{throughput}(i,k)
          \cdot f_\mathrm{fib\_available}(i,k) \cdot f_\mathrm{efficiency}(t,i) ,
          \label{eq:praw}
\end{align}
where $c_\mathrm{tar}(t,i)$ is a normalising constant that depends on the target $t$ and field $i$ (see Eq.~\eqref{eq:pnorm}). We note that $\mathbbm{1}\!\left\{\mathrm{Res}_{\mathrm{tar}=t} = \mathrm{Res}_{\mathrm{fib}=k}\right\}$ equals one if the resolution (HR or LR) of the fibre and target is the same, and it is zero otherwise. The factors $f_{\ldots}(\cdot)$ in Eq.~\eqref{eq:praw} are defined as follows.

\vspace{1mm}  The fibre tilt angle $\alpha_\mathrm{tilt}(\mathrm{tar},\mathrm{field},\mathrm{fib})$ between a target and a fibre for a given field are taken into account by $f_\mathrm{tilt}(\mathrm{tar},\mathrm{field},\mathrm{fib})$, which is non-zero for fibres that can reach the given target. In the current paper, we use the following simple definition:
\begin{equation}
        f_\mathrm{tilt}(\mathrm{tar},\mathrm{field},\mathrm{fib})=
        \begin{cases}
                1.0 & \mathrm{if} \quad \alpha_\mathrm{tilt}(\mathrm{tar},\mathrm{field},\mathrm{fib}) < \alpha_\mathrm{tilt\_max} \\
                0.0 & \mathrm{otherwise}
        \end{cases} ,
\end{equation}
where $\alpha_\mathrm{tilt\_max}$ is the maximum tilt angle of a fibre. This guarantees that fibres which cannot reach a given target are not considered. In principle, any function depending on $\alpha_\mathrm{tilt}(\mathrm{tar},\mathrm{field},\mathrm{fib}), $ in which $\alpha_\mathrm{tilt}(\mathrm{tar},\mathrm{field},\mathrm{fib}) < \alpha_\mathrm{tilt\_max}$, can be used instead of a constant value of 1.0.

\begin{figure}
\centering
\includegraphics[width=88mm]{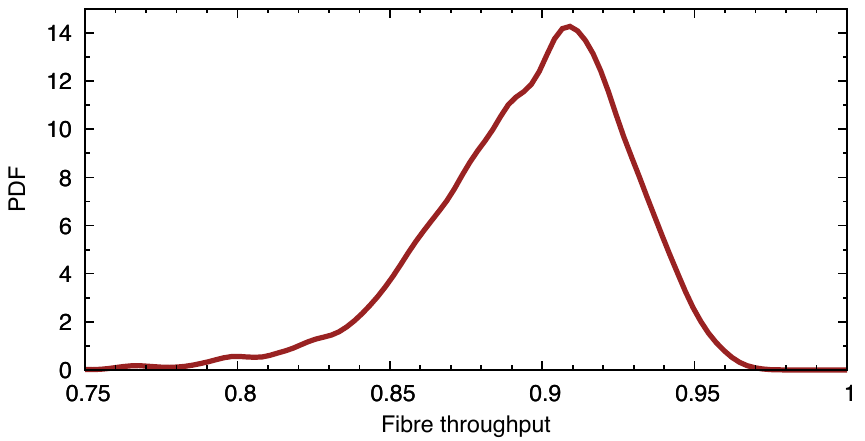}
        \caption{Distribution of fibres throughputs used in the current paper. Each fibre has a fixed throughput value during our test simulations.}
        \label{fig:fib_throughput}
\end{figure}

\vspace{1mm}It is important to note that $f_\mathrm{throughput}(\mathrm{field},\mathrm{fib})$ allows us to take the fibre throughput into account. If a target can be accessed by many fibres, a higher probability is assigned to high-throughput fibres. In cases where the throughput of each fibre is not known, this term can be ignored and $f_\mathrm{throughput}(\mathrm{field},\mathrm{fib})=1.0$. The distribution of fibre throughputs used in the current paper are shown in Fig.~\ref{fig:fib_throughput}.

\vspace{1mm}We note that $f_\mathrm{fib\_available}(i,k)$ defines a probability that a fibre $k$ in a field $i$ is available for new or repeated (see Sect.~\ref{sec:fibtotar}) science targets. Some of the fibres are used for standard stars and therefore are not available to be assigned to science targets. As a first approximation, $f_\mathrm{fib\_available}(i,k)=1.0$. For more accurate treatment, $f_\mathrm{fib\_available}(i,k)$ can be estimated by repeating the targeting algorithm hundreds or thousands of times.

\vspace{1mm} The efficiency coefficient that a given target should be observed within field $i$ is described by $f_\mathrm{efficiency}(t,i)$. This is only relevant if exposure times or sky conditions are not the same for all fields that cover a given target. For example, if a target requires a short exposure time and could be observed using short or long exposures, we can assign a very low probability to the target for fields with long exposures. Furthermore, if a target requires a long exposure during dark time and if some possible fields that include this target are observed during bright time with short exposures, then we can assign zero probability to the target in fields that are observed during bright time.

Finally, the normalising constant $c_\mathrm{tar}(t,i)$ for a target $t$ in field $i$ is determined by meeting the following condition:
\begin{align}
        &\sum\limits_{k=1}^{N_\mathrm{fib}}p_{\mathrm{tar}=t, \mathrm{field}=i, \mathrm{fib}=k} \, + \nonumber \\
        & \sum\limits_{j=1,\, j\neq i}^{N_\mathrm{fields}}\sum\limits_{k=1}^{N_\mathrm{fib}} 
        \mathbbm{1}\!\left\{ T_\mathrm{exp}(t) \leq T_\mathrm{rem}(t,j,i) \right\} 
         p_{\mathrm{tar}=t, \mathrm{field}=j, \mathrm{fib}=k} = 1.0 ,
         \label{eq:pnorm}
\end{align}
where $N_\mathrm{fib}$ is the number of fibres in one field and $N_\mathrm{fields}$ is the total number of fields that should be observed during the entire survey.
The condition $T_\mathrm{exp}(t) \leq T_\mathrm{rem}(t,j,i)$ in Eq.~\eqref{eq:pnorm} should be interpreted as a condition that the remaining exposure time is sufficient to successfully observe the target $t$. The calculation of the remaining exposure time depends on the used field order if fields have different exposure times.
In Eq.~\eqref{eq:pnorm}, $T_\mathrm{exp}(t)$ is the required exposure time of a target $t$ and $T_\mathrm{rem}(t,j,i)$ is the remaining exposure time that can be used for this target, which is summed over fields that cover the target $t$. Additionally, $T_\mathrm{rem}(t,j,i)$ is defined as
\begin{align}
        &T_\mathrm{rem}(t,j,i) = \nonumber \\ &\sum\limits_{l=j,\, l\neq i}^{N_\mathrm{fields}} 
        \mathbbm{1}\!\left\{ \exists p^\mathrm{raw}_{\mathrm{tar}=t, \mathrm{field}=l, \mathrm{fib}=k} > 0 : k=1\dots N_\mathrm{fib} \right\} 
        T_\mathrm{exp,field}(l) ,
        \label{eq:tremaining}
\end{align}
where $T_\mathrm{exp,field}$ is the exposure time for a given field. The calculation of $T_\mathrm{rem}$ assumes that the required exposure time of a target is fixed.

The probability $p$ defines the probability that an unobserved target is observed with a given fibre. Additionally, it is assumed that we start observing a target if there is sufficient time left in order to successfully complete the target. If there is not enough time left to successfully observe the target, it is removed from the target list before the probabilistic target selection (see Sect.~\ref{sec:fibtotar}).

Regarding the normalisation constant $c_\mathrm{tar}(t,i)$ in Eq.~\eqref{eq:pnorm} (see also Eq.~\eqref{eq:ptar_eq1}), the calculation of remaining time in Eq.~\eqref{eq:tremaining} assumes that we know in which order the fields are observed. The normalisation constant $c_\mathrm{tar}(t,i)$ is calculated beforehand and does not depend on the field assignment during real observations. If the field order is not known and exposure times differ in different fields, then the normalisation depends on the assumed field order. In this case, we can use a random field order to calculate the normalisation or we can perform several calculations of normalisation over many different field orders and use the mean normalising constant.

In Eq.~\eqref{eq:pnorm} the sum is performed over all fields and all fibres. The sum is undoubtedly over all fibres that could be used to start observing a given target. The defined probability only applies to new targets that have not been observed previously. Targets that require repeated observations are targeted with the highest priority. Additionally, if an observation of a target cannot be successfully completed, as the remaining exposure time is not sufficient, then the object is only targeted if no other targets are available (see Sect.~\ref{sec:fibtotar}).

The most important aspect of the normalisation given with Eq.~\eqref{eq:pnorm} is that it gives a higher probability for targets that require long exposures. In general, the normalisation counts the number of fields where a target can be observed for the first time. One aim of this normalisation is to have nearly uniform completeness that does not depend on exposure time or magnitude. This aspect is later analysed in Sect.~\ref{sec:selfunc} and Fig.~\ref{fig:example_texp_completeness}.

\subsection{Probabilistic targeting algorithm}
\label{sec:targeting_algorithm}

A target is assigned to each fibre in the targeting algorithm. For a given fibre $k$ in a field $i$, we assigned a probability to each target that can be targeted. This only applies for targets that are not yet observed. Targets that need repeated observations are allocated prior to this (see Sect.~\ref{sec:fibtotar}). The probability that a target $t$ should be assigned to a fibre $k$ in a field $i$ is defined as
\begin{align}
        p^\mathrm{tar}_{\mathrm{tar}=t, \mathrm{field}=i, \mathrm{fib}=k} = \,& c_\mathrm{fib}(i,k) \cdot
        p_{\mathrm{tar}=t, \mathrm{field}=i, \mathrm{fib}=k} \cdot \nonumber \\
        & f_\mathrm{compl}(t) \cdot F_\mathrm{survey}(i,s:t\in s) ,
        \label{eq:ptar}
\end{align}
where $f_\mathrm{compl}$ is a small-scale-merit function that defines the fraction of targets averaged over a small sky area that should be observed for a successful survey\footnote{The 4MOST survey consist of several sub-surveys. From here on, $s$ represents a single sub-survey.}. If a survey is expected to observe a constant fraction of targets independent of target parameters and coordinates, then $f_\mathrm{compl}(t\in s)=\mathrm{const} \leq 1.0$ for a survey $s$.

The factor $F_\mathrm{survey}(i,s)$ takes into account that the number density of targets varies between surveys, that targets are clustered, and that there is a fixed fibre pattern in the 4MOST instrument (see Fig.~\ref{fig:fibres_multi}). The factor $F_\mathrm{survey}(i,s)$ is defined as
\begin{equation}
        F_\mathrm{survey}(i,s) = \frac{1}{N^\mathrm{sur}_\mathrm{fib\_acc}(i,s)} \cdot
        n^\mathrm{sur}_\mathrm{tar\_per\_fib}(i,s) \cdot \frac{f_\mathrm{add}(s)}{f^\mathrm{sur}_\mathrm{suc}(i,s)} ,
        \label{eq:Fsurvey}
\end{equation}
where each component is described below. We note that $F_\mathrm{survey}(i,s)$ is calculated per field, based on targets that are accessible by fibres in field $i$.

The probabilities $p_{\mathrm{tar}=t, \mathrm{field}=i, \mathrm{fib}=k}$ do not take the fixed fibre pattern of the 4MOST instrument into account. Additionally, the probabilities $p_{\mathrm{tar}=t, \mathrm{field}=i, \mathrm{fib}=k}$ assume that all fibres in a field can be used for all targets in a single survey. These assumptions are not valid in reality and are taken into account by $F_\mathrm{survey}(i,s)$. The parameter $N^\mathrm{sur}_\mathrm{fib\_acc}(i,s)$ takes into account that not all fibres are accessible by all surveys by counting the number of fibres in a field $i$ that are accessible by a survey $s:$
\begin{equation}
        N^\mathrm{sur}_\mathrm{fib\_acc}(i,s) = \sum\limits_{k=1}^{N_\mathrm{fib}}
        \mathbbm{1}\!\left\{ \exists p_{\mathrm{tar}=t, \mathrm{field}=i, \mathrm{fib}=k} > 0 : t \in s \right\} .
\end{equation}
The parameter $n^\mathrm{sur}_\mathrm{tar\_per\_fib}(i,s)$ determines the mean number of targets (from all surveys) per fibre for the fibres that are accessible by survey $s$. It is defined as
\begin{align}
        &n^\mathrm{sur}_\mathrm{tar\_per\_fib}(i,s) = \frac{1}{N^\mathrm{sur}_\mathrm{fib\_acc}(i,s)} \cdot \nonumber \\
        &\qquad\sum\limits_{k=1}^{N_\mathrm{fib}}
        \mathbbm{1}\!\left\{ \exists p_{\mathrm{tar}=t, \mathrm{field}=i, \mathrm{fib}=k} > 0 : t \in s \right\}
        n_\mathrm{tar\_per\_fib}(i,k) ,
\end{align}
where $n_\mathrm{tar\_per\_fib}(i,k)$ is the number of potential targets per fibre
\begin{equation}
        n_\mathrm{tar\_per\_fib}(i,k) = \sum\limits_{t=1}^{N^\mathrm{all}_\mathrm{tar}}
        \mathbbm{1}\!\left\{ p_{\mathrm{tar}=t, \mathrm{field}=i, \mathrm{fib}=k} > 0 \right\} ,
\end{equation}
where $N^\mathrm{all}_\mathrm{tar}$ represents the number of all targets over all surveys.

\begin{figure}
\centering
\includegraphics[width=88mm]{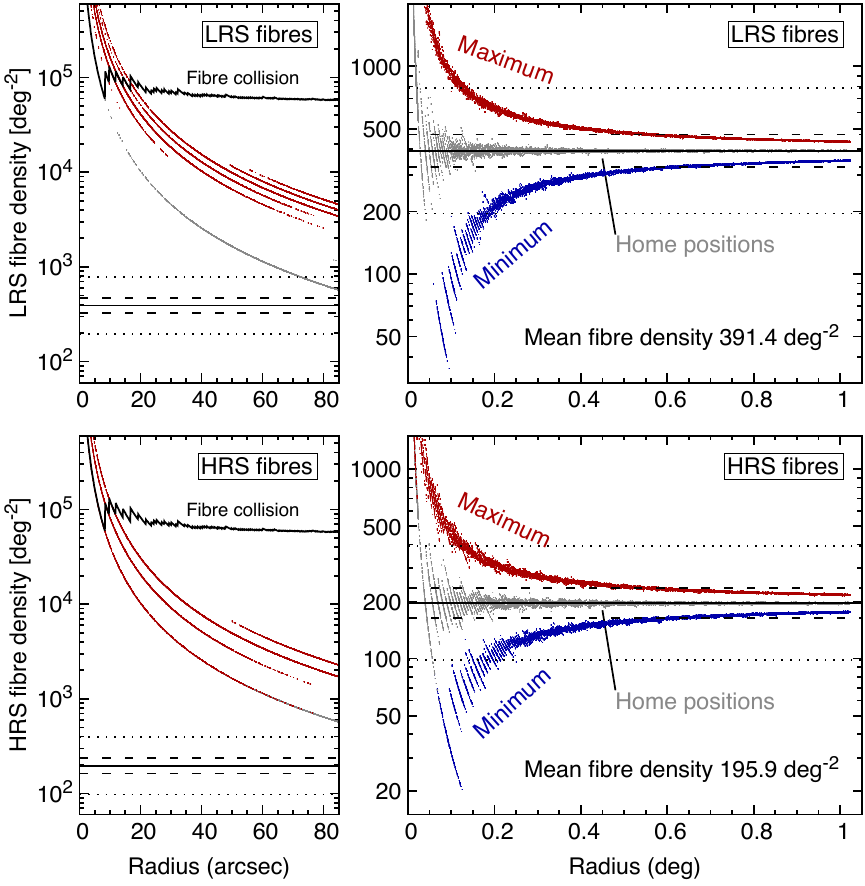}
        \caption{Fibre density for low-resolution (upper panel) and high-resolution (lower panel) spectrograph fibres. Right-hand panels show the fibre density up to one~degree, while left-hand panels show the fibre density at small angular scales. Fibre density is defined inside a circle (circumscribed the field of view) as a function of radius (grey points). Fibre density depends on circle location due to the fixed fibre pattern (see Fig.~\ref{fig:fibres_multi}) and the scatter is higher when the radius is smaller. Red and blue points show the maximum and minimum fibre number density inside a circle, taking the fibre patrol radius into account. The solid black line shows the mean fibre density in the 4MOST field of view, which is 391.4 and 195.9 fibres per square degree for low- and high-resolution spectrograph fibres, respectively. The black dashed line shows 1.2 times higher or lower density, and the black dotted line shows two times higher or lower density compared with fibre mean density. In a circle with a radius of 0.2 degrees, the fibre density can be twice as high or low as the mean value. In the left-hand panels, the black solid line shows fibre collision regions, where everything above the line is forbidden due to fibre collisions.}
        \label{fig:fib_density}
\end{figure}

\begin{figure}
\centering
\includegraphics[width=88mm]{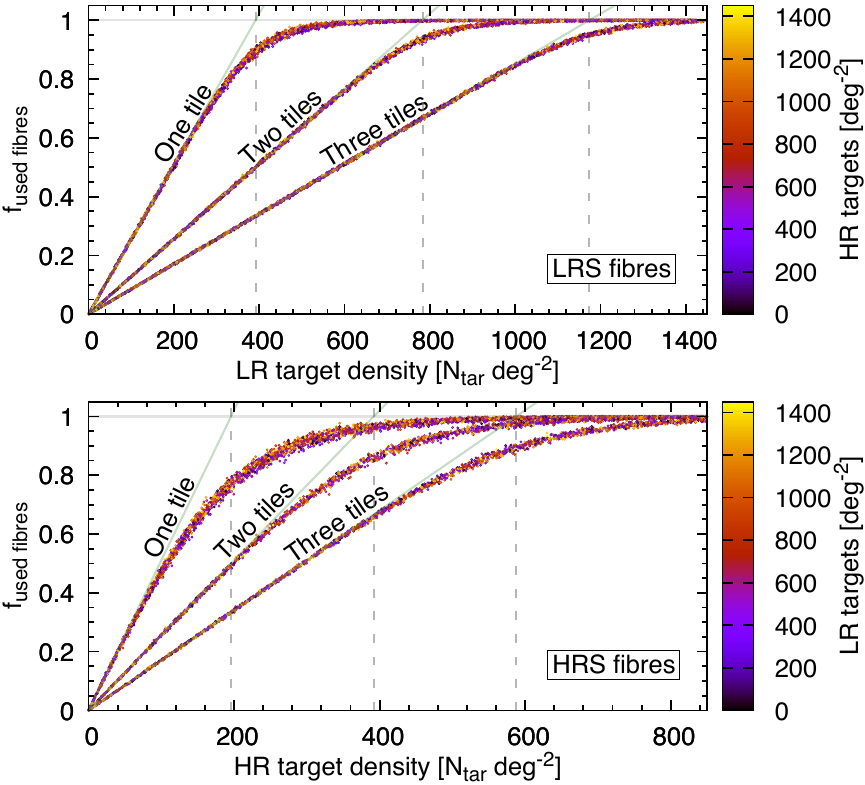}
        \caption{Fraction of used fibres as a function of target density, assuming a Poisson distribution of targets in the sky. The \emph{upper} and \emph{lower} panels show fibre usage efficiency for low- and high-resolution spectrograph fibres, respectively. Colours indicate target density for high- (upper panel) and low-resolution (lower panel) targets. The targeting simulation has been run using one, two, or three tiles (pointings) that are overlaid with small random shifts, which are indicated by three different groups of points on both panels. Each target was observed only once and all fibres were used for science targets. Solid green lines show theoretical expectations for perfect targeting. Vertical dashed lines indicate the fibre densities on the sky using one, two, or three overlapping tiles.}
        \label{fig:poisson_test_fib}
\end{figure}

\begin{figure}
\centering
\includegraphics[width=88mm]{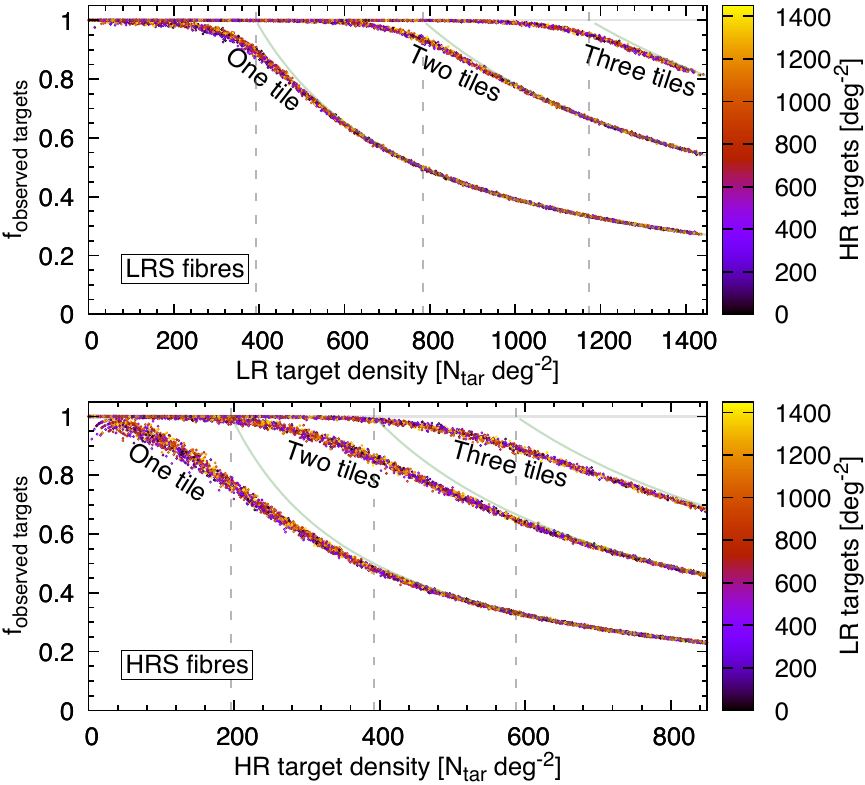}
        \caption{Fraction of observed targets as a function of target density for low- (upper panel) and high-resolution (lower panel) targets. The targeting simulation has been run using one, two, or three tiles overlaid (with small random shifts). Each target was observed only once and all fibres were used for science targets. Green solid lines show theoretical expectations for perfect fibre to target allocation. Vertical dashed lines show fibre density on the sky for low- (upper panel) and high-resolution (lower panel) spectrograph fibres.}
        \label{fig:poisson_test_tar}
\end{figure}

The factor $N^\mathrm{sur}_\mathrm{fib\_acc}(i,s)$ takes into account that not all fibres are accessible by all surveys.\ Also, for surveys with sparse sampling of targets, which have a lower $N^\mathrm{sur}_\mathrm{fib\_acc}(i,s)$ value, the targeting probability is increased. Additionally, we have to take into account that target density varies within a field of view, hence some fibres can reach more targets than other fibres. This is described by $n^\mathrm{sur}_\mathrm{tar\_per\_fib}(i,s)$. If a survey can only use fibres that can reach more targets than average, then the targeting probability for this survey's targets is increased.

\begin{figure*}
\centering
\includegraphics[width=180mm]{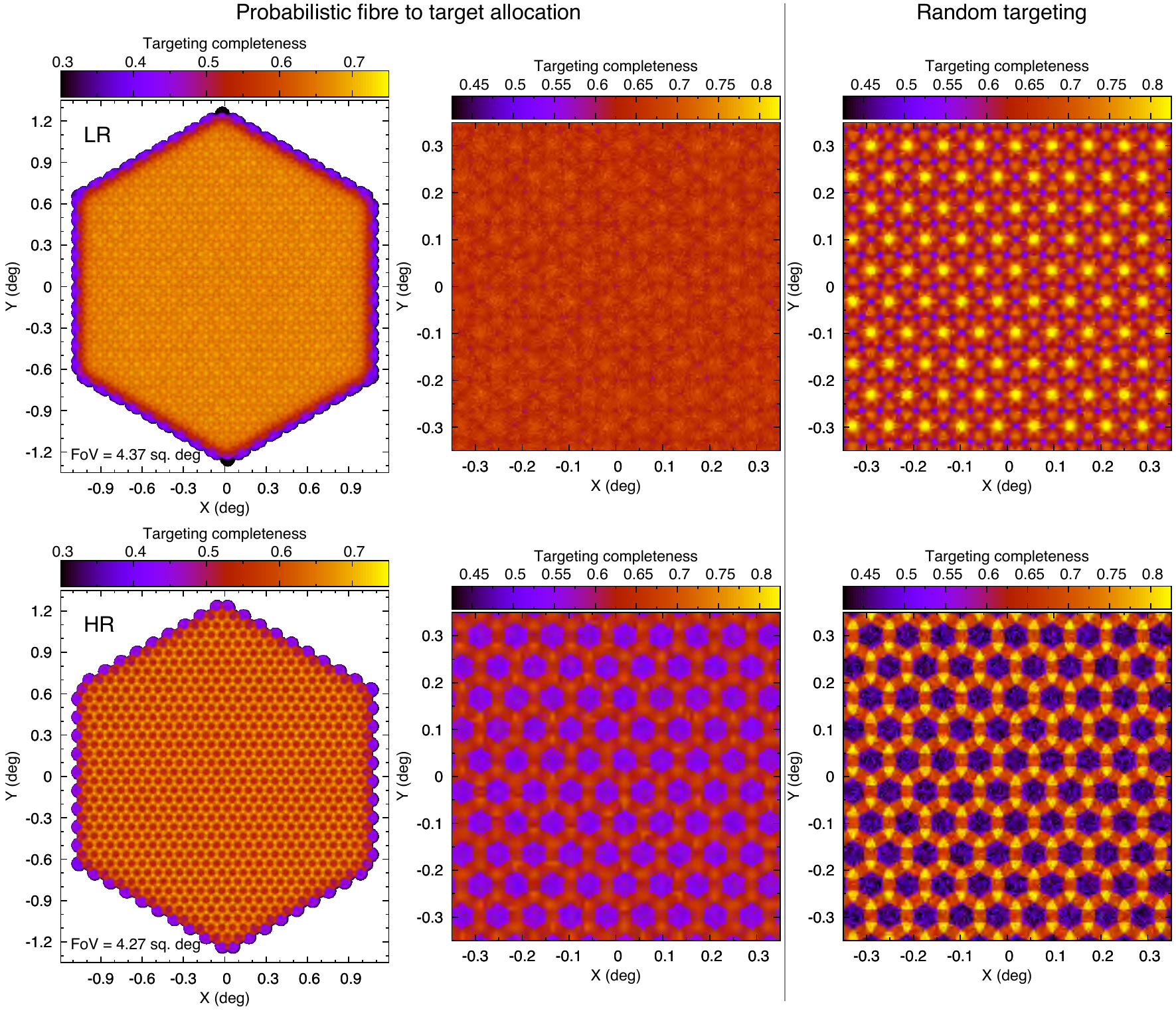}
        \caption{Targeting completeness in a single field of view. The top row shows the completeness map for low-resolution targets and the bottom row for high-resolution targets. In the targeting simulation, we used Poisson distributed targets across the sky with 1.5 times higher number density than the number density of fibres. Targeting was performed using the probabilistic fibre-to-target assignment as described in Sect.~\ref{sec:target_allocation}. Left panels show the full field of view for a single field. The completeness decreases towards the edge of the field. Middle panels show a zoom-in region of the left panel. For comparison, in the right panels we show a completeness map for random targeting in which each fibre-target pair has the same probability. In the right panel, we can clearly see the patrol regions of fibres (see also Fig.~\ref{fig:fibres_multi}), which are significantly reduced when using probabilistic fibre-to-target assignment.}
        \label{fig:fov_completeness}
\end{figure*}

\begin{figure*}
\centering
\includegraphics[width=180mm]{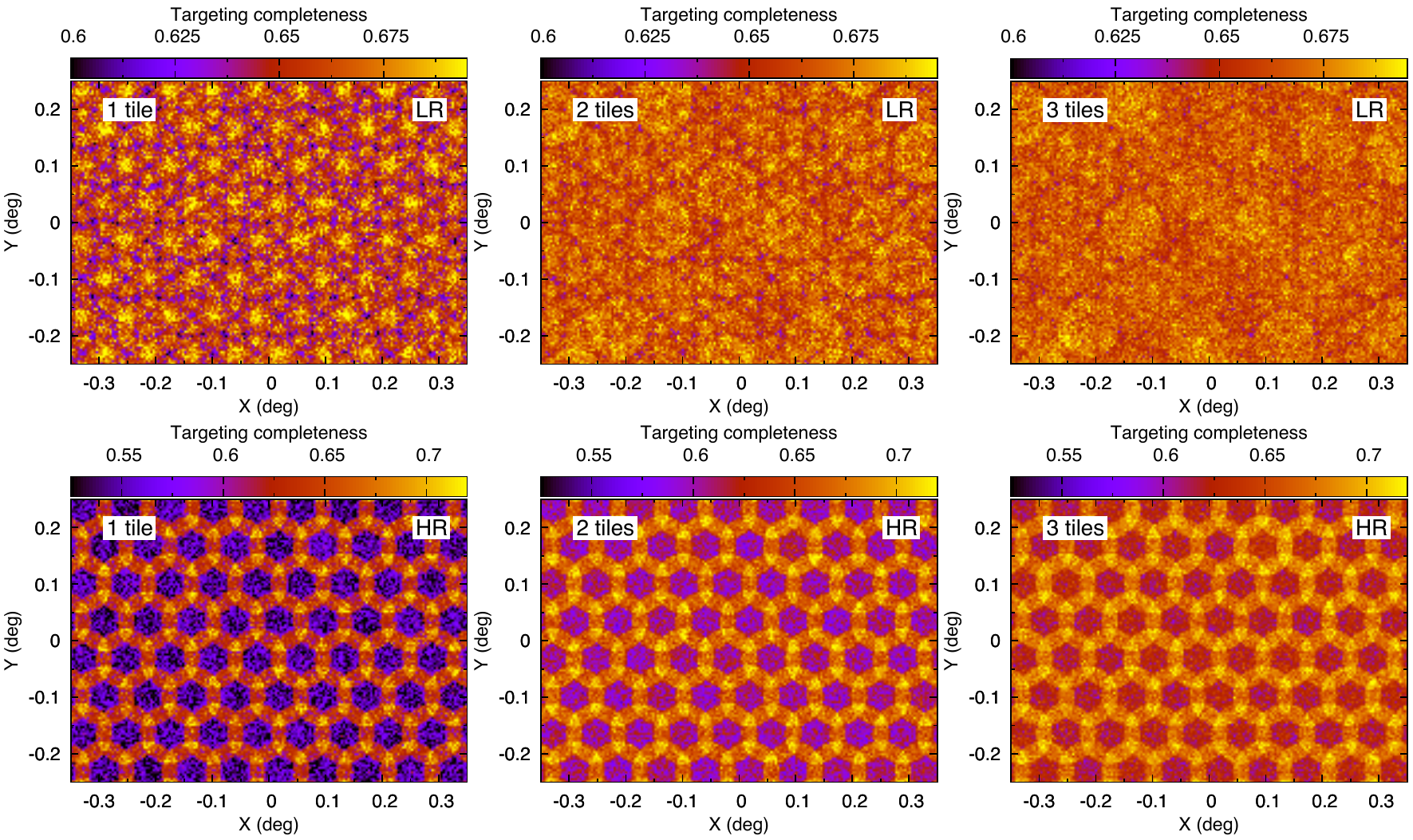}
        \caption{Targeting completeness for Poisson distributed targets using one, two, or three visits. The number density of targets was set to 1.5 times higher than the total number density of fibres. For low-resolution spectrograph fibres, the patrol area of some fibres is visible and they have slightly higher completeness. This reflects the varying throughput of fibres. Probabilistic targeting slightly prefers fibres with a higher throughput. However, this effect is only visible if the field is observed more than once. For a single visit, a fixed fibre pattern determines the completeness on small angular scales.}
        \label{fig:fov_completeness_multi}
\end{figure*}

An additional factor $f^\mathrm{sur}_\mathrm{suc}(i,s)$ is added to the parameter $F_\mathrm{survey}(i,s)$ to better balance the fibre usage between surveys. This factor is estimated based upon targeting simulations. Based on these simulations, for each target $t$ we can estimate the probability $p^\mathrm{tar}_\mathrm{success}$ that it was successfully observed with sufficient exposure time. The factor $f^\mathrm{sur}_\mathrm{suc}(i,s)$ is defined as
\begin{equation}
        f^\mathrm{sur}_\mathrm{suc}(i,s) \!=\! \frac{\sum\limits_{t \in s}
        \!\mathbbm{1}\!\left\{ \exists p_{\mathrm{tar}=t, \mathrm{field}=i, \mathrm{fib}=k} \!>\! 0 \!:\! k\!=\!1\dots N_\mathrm{fib} \right\} p^\mathrm{tar}_\mathrm{success}(t) }
        { \sum\limits_{t \in s}
        \!\mathbbm{1}\!\left\{ \exists p_{\mathrm{tar}=t, \mathrm{field}=i, \mathrm{fib}=k} \!>\! 0 \!:\! k\!=\!1\dots N_\mathrm{fib} \right\} f_\mathrm{compl}(t) }.\!
\end{equation}
This factor is calculated per field and it boosts surveys that do not observe enough targets compared to their expected number of targets ($f_\mathrm{compl}$). Initially we can assume $f^\mathrm{sur}_\mathrm{suc}(i,s)=1.0$.

Furthermore, we can manually add another factor $f_\mathrm{add}(s)$  if none of the previous adjustments present satisfactory results. For the majority of surveys, this factor should be $f_\mathrm{add}(s)=1.0$.

The probabilities $p^\mathrm{tar}_{\mathrm{tar}=t, \mathrm{field}=i, \mathrm{fib}=k}$ without a normalising constant $c_\mathrm{fib}(i,k)$ are calculated beforehand for all fibre-target pairs.  The normalising constant in Eq.~\eqref{eq:ptar} normalises the targeting probabilities for each fibre, and it is calculated during fibre to target assignment as explained in Sect.~\ref{sec:fibtotar}. The normalisation cannot be calculated beforehand as it depends on the set of targets that are available, including fibre collisions, for a given fibre.

\subsection{Fibre to target assignment}
\label{sec:fibtotar}

During fibre to target assignment, the aim is to find a target for each fibre, to reserve some fibres for standard stars, and to allocate sky fibres. In the current paper, we use a simple and straightforward algorithm for this. It requires additional analysis to find the most optimal algorithm. In our test simulations, we used the following algorithm for targeting:
\begin{enumerate}
        \item Fibres are allocated to standard stars. If the number of good standard stars in a field of view is large enough, then it is preferential to use fibres that cannot be allocated to science targets.
        \item Fibres are allocated to targets that require repeated observations. The highest priority is to allocate fibres to targets that have been previously observed but need more exposure time to be completed. This guarantees that most of the observed targets are successfully completed. Wherever possible, fibres should be used that cannot be allocated to new science targets, increasing the efficiency of fibre to target allocation. Additionally, fibres that are closer to the target and/or fibres with fewer potential new targets are preferred.
        \item Targets that cannot be observed due to fibre collisions are removed. All targets that are closer than the fibre collision radius for previously allocated fibres need to be removed.
        \item Targets that require more time than is remaining are removed. These targets are put on hold and are then observed if some fibres are left empty.
        \item New science targets are allocated to each fibre as follows:
        \begin{enumerate}
                \item An empty fibre, which is not yet allocated, needs to be found and can then be used for a new science target. This can be performed randomly or by ranking the fibres by the number of remaining targets per fibre and selecting the one with the lowest number of targets per fibre. The latter is more efficient because it first uses fibres with only one or a few targets.
                \item A target for a fibre using probabilities $p^\mathrm{tar}_{\mathrm{tar}=t, \mathrm{field}=i, \mathrm{fib}=k}$, as described below, must be found.
                \item Targets around a recently allocated target, which cannot be observed due to fibre collisions, need to be removed.
                \item The step from point (a) needs to be repeated and a new empty fibre must be found. However, if there are no empty fibres left or if the science target list is empty, the search for empty fibres should stop.
        \end{enumerate}
        \item Sky fibres need to be allocated. First, fibres that are not yet allocated should be used. If the number density of empty fibres is not high enough, some science fibres should be reallocated as sky fibres. Then fibres that have been allocated to new science targets should be preferentially selected.
        \item The remaining empty fibres to science targets that were put on hold, as they require more time than is remaining, need to be allocated. Our assumption is that even partially observed targets are scientifically useful. If this is not the case for some targets, then these targets should be removed completely from the potential target list. Additionally, targets that can reach higher completion are preferred if there is more than one target available per fibre.
        \item The remaining fibres are allocated to auxiliary science targets, that is, targets that are not part of survey's input catalogues but might be scientifically interesting per se.
\end{enumerate}

To find a new science target $t$ for an empty fibre $k$ in a field $i$, we can use the probabilities $p^\mathrm{tar}_{\mathrm{tar}=t, \mathrm{field}=i, \mathrm{fib}=k}$. At this point only targets that are as follows should be considered:
\begin{itemize}
        \item have not been previously observed;
        \item require less exposure time than remains for this target (i.e. the sum of exposure times of all remaining fields in a target's location);
        \item can be reached by a fibre $k$, that is, a target is within a patrol radius of a fibre and targets in fibre collision regions are not considered.
\end{itemize}
The set of these targets is designated as $\mathbf{t}_\mathrm{accessible}(i,k)$ and the probabilities $p^\mathrm{tar}_{\mathrm{tar}=t, \mathrm{field}=i, \mathrm{fib}=k}$ are normalised so as to satisfy
\begin{equation}
        \sum\limits_{t \in \mathbf{t}_{\mathrm{accessible}(i,k)} } p^\mathrm{tar}_{\mathrm{tar}=t, \mathrm{field}=i, \mathrm{fib}=k} = 1.0.
\end{equation}
Once we have the normalised probabilities $p^\mathrm{tar}_{\mathrm{tar}=t, \mathrm{field}=i, \mathrm{fib}=k}$ for all targets accessible by a given fibre, we can select one of them at random according to the probabilities assigned.

\section{Target allocation for Poisson distributed targets}
\label{sec:poisson_test}

In this section we run some basic tests using Poisson distributed targets. Random targets in the sky are the simplest case and this analysis gives a rough estimate of what we should expect from the targeting algorithm with a given number density of targets. A more realistic distribution of targets, such as clustered targets, are analysed in Sect.~\ref{sec:example_test}.

As a starting point, we must analyse the 4MOST field of view. Figure~\ref{fig:fibres_multi} shows the fibre pattern layout over a field of view for low- and high-resolution spectrograph fibres. The fibre density in the 4MOST field of view is not uniform across small scales. Table~\ref{table:fib_coverage} gives the fraction of area that is covered by one to six fibres for low- and high-resolution spectrograph fibres. In the low-resolution regime, most of the field of view can be reached by three or four fibres. In the high-resolution regime most of the field can only be reached by one or two fibres. Hence, in high-resolution, targeting is less efficient than for low-resolution targets.

Figure~\ref{fig:fib_density} shows the fibre density across the 4MOST field of view. On average there are 391 low- and 196 high-resolution spectrograph fibres per square degree. However, due to the geometry of the fibre patrol areas, the number density can vary significantly, particularly over smaller areas. Figure~\ref{fig:fib_density} also shows the maximum and minimum fibre density, whilst taking into account the fibre patrol area. For example, in a circle with radius 0.2~degrees, the number density of fibres can be up to twice as high or low as the mean fibre density; in a circle with a radius of 0.6~degrees, the number density can be up to 20 percent higher or lower than the mean fibre density. 

Physically, we cannot place two fibres closer than 17~arcsec, which limits the maximum fibre density over a small area. The fibre collision line is shown on the left-hand panels in Fig.~\ref{fig:fib_density}. The fibre collision is only a dominant problem in an area with a radius less than 0.5~arcmin.

For simple test simulations, we generated Poisson distributed targets across the sky with different number densities. We generated different targets with differing number densities for low- and high-resolution. For the test simulation, we only used one 4MOST field with either one, two, or three tiles (pointings) stacked on top of each other. In the case of two or three tiles, we slightly shifted the field centres in order to mitigate the fixed fibre pattern. The targeting algorithm was run as described in Sect.~\ref{sec:fibtotar}. Each target was observed only once (no repeat observations) and all fibres were used for science targets (zero fibres were allocated to standard stars and sky).

Figures~\ref{fig:poisson_test_fib} and~\ref{fig:poisson_test_tar} show the results of the test simulations using Poisson distributed targets. Figure~\ref{fig:poisson_test_fib} shows the fraction of used fibres as a function of target density. If the target density is low, then nearly all targets are observed and the fraction of used fibres increases linearly. If the number density of targets is significantly higher than the fibre density, all fibres are successfully used for science targets. Vertical lines in Fig.~\ref{fig:poisson_test_fib} show the fibre density multiplied by one, two, or three. If the target density is equal to the fibre density, then we can use $\sim$90\,\% of low-resolution and $\sim$75\,\% of high-resolution spectrograph fibres when using only one pointing. If we observe the same region more than once, then the fraction of used fibres increases due to increased flexibility of target allocation. During real observations, the fibres that cannot be used for science targets are then used as sky fibres.

Figure~\ref{fig:poisson_test_tar} shows the fraction of observed targets as a function of target density. If the number density of targets is low, then most of the targets are observed. If the target density is significantly higher than the fibre density, then the fraction of observed targets follows the theoretical expectation, which is that  fibre density is divided by target density.

\begin{figure}
\centering
\includegraphics[width=88mm]{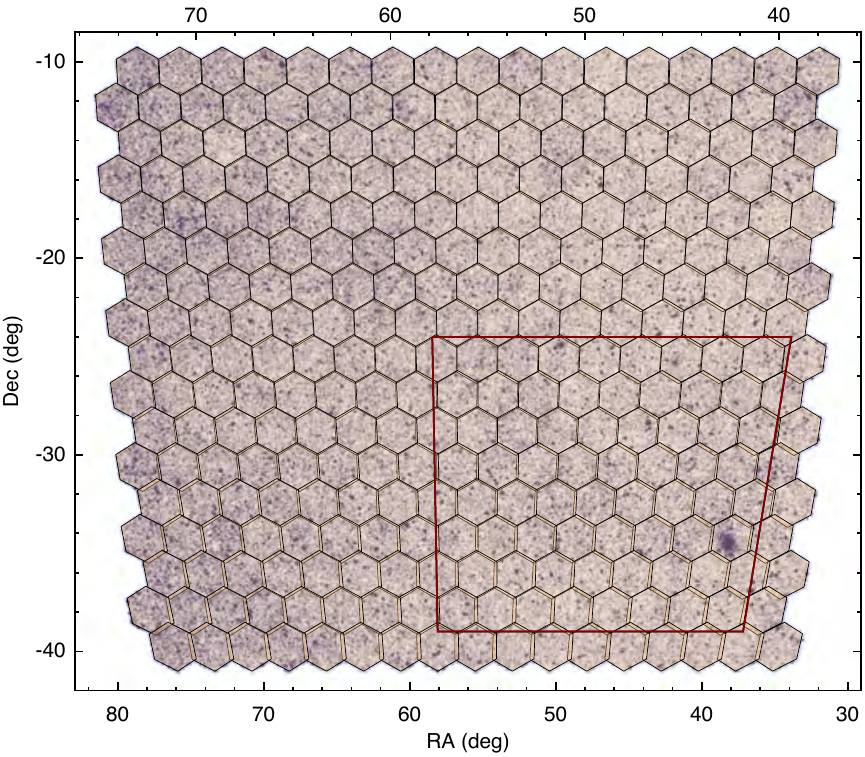}
        \caption{Tiling pattern across the sky area used during our tests. The underlying heat map shows the number density of targets on the sky. Darker areas show higher number density of targets. The light gradient from left to right is because of the Milky Way stellar surveys; there are more stars close to the Milky Way disc, which is located on the left side of the selected survey area. The red box highlights the smaller area that we use as a zoom region on Figs.~\ref{fig:targets}, \ref{fig:example_map_completeness}, and \ref{fig:example_map_success}. }
        \label{fig:tiles}
\end{figure}

Since we observed both low- and high-resolution targets simultaneously, the obvious question to explore is how the fraction of used fibres or fraction of observed targets depends on the target density of alternative resolution. In Figs.~\ref{fig:poisson_test_fib} and~\ref{fig:poisson_test_tar} the colour of the points show the target density of alternative resolution. These figures show that there is no clear dependence on the target density of alternative resolution, that is, the fraction of observed low-resolution targets does not depend on the target density of high-resolution spectrograph fibres, and vice versa. The scatter seen in Figs.~\ref{fig:poisson_test_fib} and~\ref{fig:poisson_test_tar} is caused by the Poisson distributed nature of the generated targets.

\begin{figure*}
\centering
\includegraphics[width=180mm]{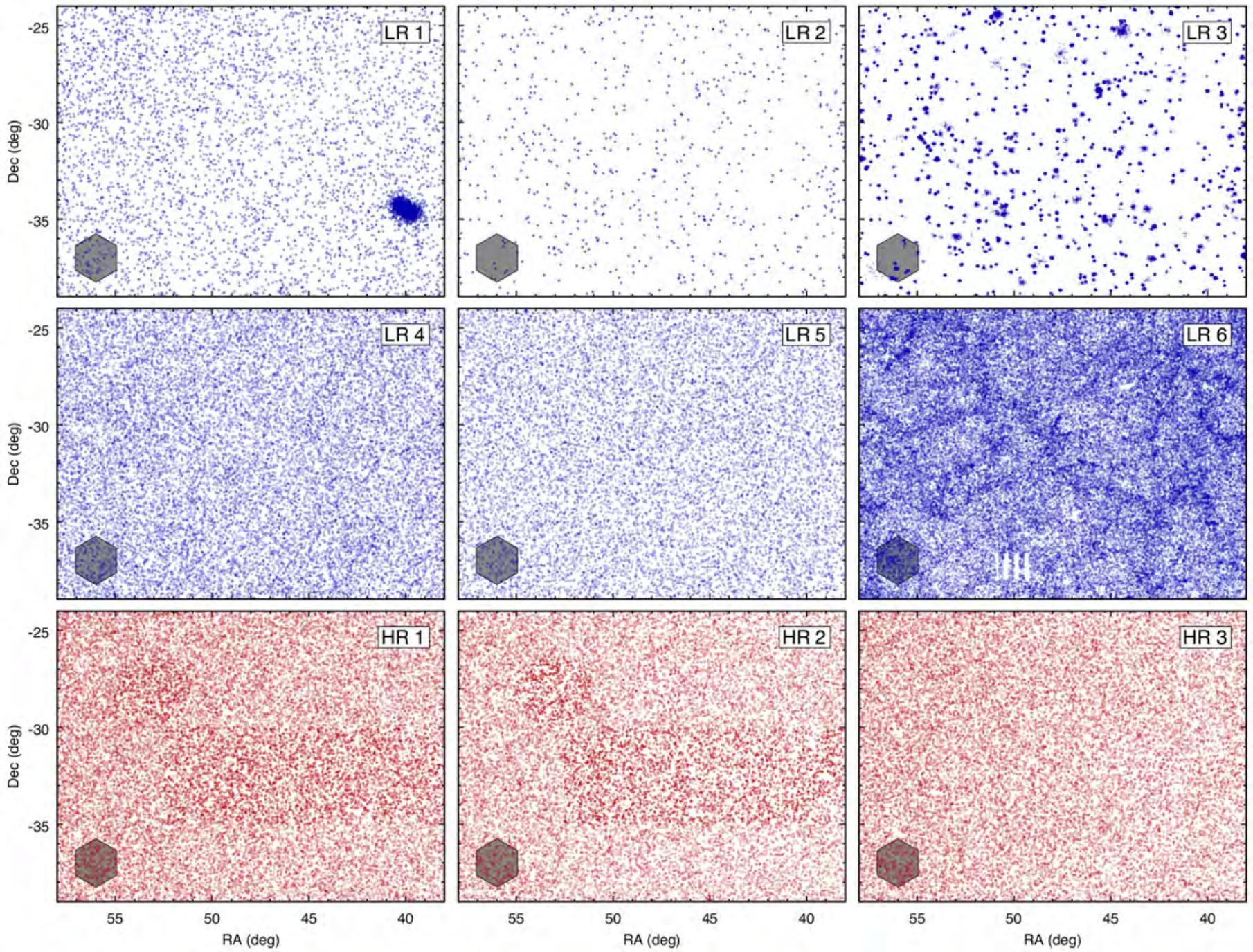}
        \caption{Sky coordinates of targets in six low-resolution and three high-resolution surveys used during our test. A telescope field of view is superimposed.}
        \label{fig:targets}
\end{figure*}

Another question we consider is the targeting completeness across the field of view and how it is affected by the fixed fibre pattern. To test this, we ran a test simulation with 1.5 times higher number density of targets than fibre density. We ran the targeting simulation thousands of times to accumulate sufficient statistics. The results are shown in Fig.~\ref{fig:fov_completeness}. The edges of the field in low- and high-resolution are different due to a differing fibre pattern across low- and high-resolution. Hence, the accessible field of view for low-resolution targets is 4.37~$\mathrm{deg}^2$, and for high-resolution targets it is 4.27~$\mathrm{deg}^2$. In both cases, the targeting completeness drops at the edges of the field. With regards to the fibre pattern, in low-resolution the pattern is only weakly noticeable as probabilistic fibre to target assignment has erased most of the fibre pattern. In high-resolution the fibre pattern is clearly visible as a large fraction of the field of view can only be reached by one fibre (see Table~\ref{table:fib_coverage}).

Considering the selection function, one advantage of the proposed probabilistic fibre to target assignment algorithm is that the fixed fibre pattern is significantly mitigated. To demonstrate this, we ran the targeting algorithm by selecting a random target for each fibre. The results are shown on the right-hand panels of Fig.~\ref{fig:fov_completeness}. We can clearly see that in low- and high-resolution, the fixed fibre pattern is strongly visible on the sky. Total completeness over one field is the same for both random and probabilistic targeting. The advantage of probabilistic targeting is that the fibre pattern is significantly mitigated for high-resolution or almost erased for low-resolution on the sky. Figure~\ref{fig:fov_completeness} shows the results for a single pointing. If the same region is observed more than once, the fixed fibre pattern is almost erased, even for high-resolution targets. This is shown in Fig.~\ref{fig:fov_completeness_multi}.

\section{An example test case with mock surveys}
\label{sec:example_test}

\begin{table}
\caption{Target densities and exposure times of six low-resolution and three high-resolution surveys used in this example. Target densities are estimated within a circle with a 1~degree radius. We give the median density ($\rho_{50}$) as well the 10 and 90 percent quantiles of the distribution. For some surveys the target density is uniform across the selected area whilst for others the target density varies significantly. For exposure times we give both the mean and median values.}
\label{table:surveys}
\centering
\begin{tabular}{l c c c c c c}
\hline\hline
Survey & $\rho_{50}$ & $\rho_{10}$ & $\rho_{90}$ & 
$\!\operatorname{mean}(T_\mathrm{exp})\!$ & $\!\operatorname{med}(T_\mathrm{exp})\!$ \\
\hline
   &   $\mathrm{deg}^{-2}$   & $\mathrm{deg}^{-2}$ & $\mathrm{deg}^{-2}$ & min & min  \\
\hline
        LR~1 & 22 & 14 & 50    & 90 & 90 \\
        LR~2 & 3 & 1 & 4       & 16 & 31 \\
        LR~3 & 73 & 31 & 144   & 16 & 22 \\
        LR~4 & 75 & 60 & 95    & 19 & 44 \\
        LR~5 & 51 & 47 & 56    & 38 & 53 \\
        LR~6 & 375 & 305 & 473 & 25 & 25 \\
        \textbf{LR~tot} & \textbf{618} & \textbf{529} & \textbf{737} & \textbf{25} & \textbf{33} \\
        \hline
        HR~1 & 119 & 88 & 158 & 30 & 36 \\
        HR~2 & 89 & 63 & 125  & 63 & 64 \\
        HR~3 & 99 & 72 & 151  & 87 & 76 \\
        \textbf{HR~tot} & \textbf{306} & \textbf{224} & \textbf{431} & \textbf{55} & \textbf{58} \\
\hline
\end{tabular}
\end{table}

\begin{figure}
\centering
\includegraphics[width=88mm]{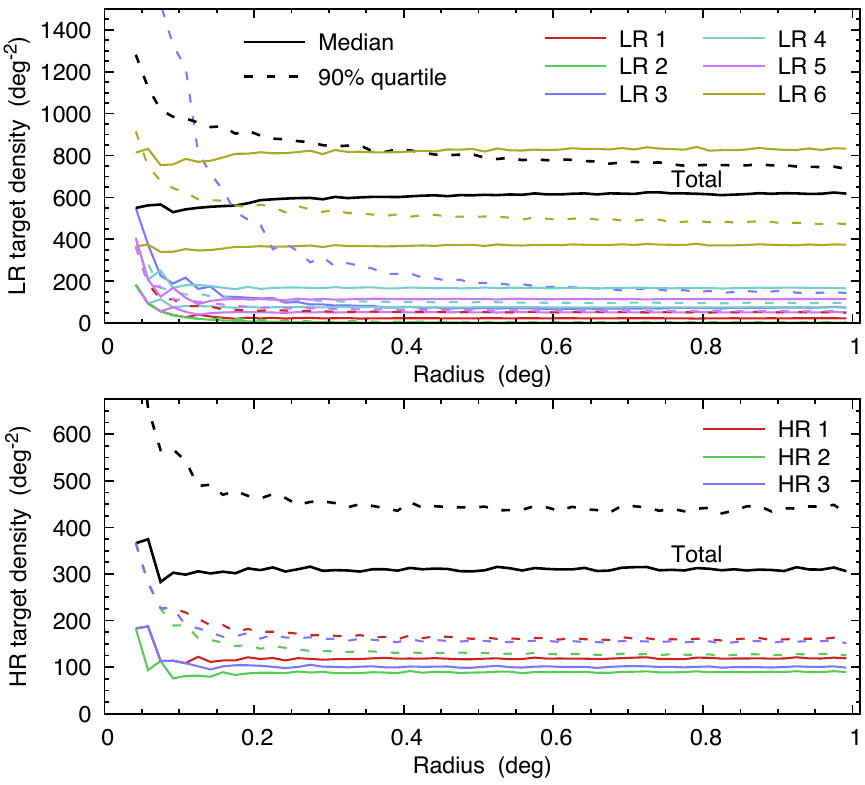}
        \caption{Target density on the sky for low- (upper panel) and high-resolution (lower panel) test surveys. Target density was calculated within a circle (with random location within a fixed sky area) of radius $r$ as indicated in abscissa. Black lines show the total target density for a given resolution, coloured lines show target density for individual surveys. Solid lines show median target density on the sky, while dashed lines indicate the 90\,\% quantiles. Due to the clustering of targets, the target density (90\,\% quantile) increases towards small sky areas.}
        \label{fig:example_tar_density}
\end{figure}

\begin{table}
\caption{Surveys in our example. Targeting parameters for surveys averaged over all fields. The definitions of these parameters are given in Sect.~\ref{sec:target_allocation}.}
\label{table:surveys_targeting}
\centering
\begin{tabular}{l c c c c}
\hline\hline
Survey & $f_\mathrm{compl}$ & $N^\mathrm{sur}_\mathrm{fib\_acc}$ & $n^\mathrm{sur}_\mathrm{tar\_per\_fib}$ & $f^\mathrm{sur}_\mathrm{suc}$ \\
\hline
        LR~1 & 0.6 & 375 & 6.8 & 1.33 \\
        LR~2 & 1.0 &  38 &25.7 & 0.82 \\
        LR~3 & 0.8 & 161 &28.3 & 0.85 \\
        LR~4 & 0.9 & 810 & 6.8 & 1.02 \\
        LR~5 & 0.1 & 587 & 6.8 & 8.10 \\
        LR~6 & 0.5 & 1514& 7.4 & 1.86 \\
        \hline
        HR~1 & 0.5 & 525 & 4.4 & 1.27 \\
        HR~2 & 0.5 & 443 & 4.7 & 0.87 \\
        HR~3 & 0.5 & 494 & 5.0 & 0.87 \\
\hline
\end{tabular}
\end{table}

\begin{figure}
\centering
\includegraphics[width=88mm]{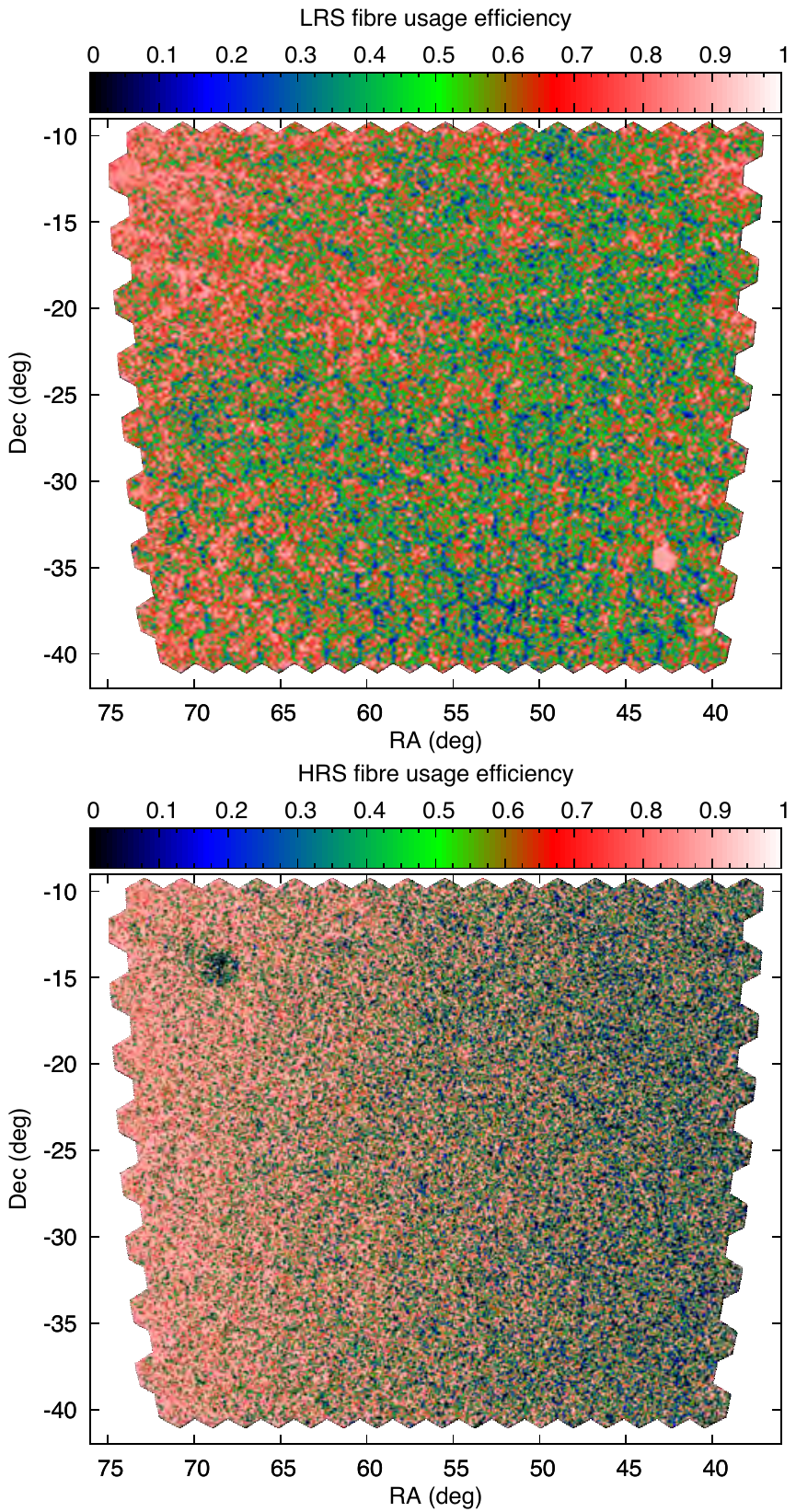}
        \caption{Fibre usage efficiency for low- (upper panel) and high-resolution (lower panel) surveys. The visible gradient from left to right is due to the fixed tiling and varying number density of objects (see Fig.~\ref{fig:tiles}).}
        \label{fig:example_fibre_efficiency}
\end{figure}

\subsection{Setup of the test simulations}

In this section we run a test, where we use three high-resolution surveys and six low-resolution surveys, which are observed simultaneously. These surveys were selected to somewhat maximise the difference between them, that is, the number density, clustering, and targets exposure times distribution.
The surveys were selected from 4MOST mock target catalogues. The high-resolution surveys and LR~1 are Milky Way stellar surveys constructed using the Galaxia model \citep{2011ApJ...730....3S}. LR~2 and LR~3 are galaxy cluster surveys, where LR~2 represents cluster main galaxies and LR~3 illustrates cluster member galaxies. The mock catalogue of the clusters is based on the MultiDark dark matter only $N$-body simulations \citep{2016MNRAS.457.4340K}. The construction of the mock is described in \citet{2019MNRAS.tmp.1335C}. LR~4 represents the eRosita AGN mock catalogue and LR~5 is the mock catalogue of TiDES host galaxies. The distribution of targets in LR~4 and LR~5 is uniform in the current mock catalogues. The last survey, LR~6, is a galaxy redshift survey constructed based on the SDSS data\footnote{LR~6 is not part of the 4MOST mock catalogues. It is used in the current paper to represent a realistic galaxy redshift survey.}. The selection of these surveys is arbitrary and in the current paper they are used to analyse the main aspects of the target allocation problem. The presented conclusions do not depend on the choice of the selected mock surveys.

\begin{table*}
\caption{Results of the targeting algorithm for five different simulations with slightly altered targeting algorithm parameters. For each survey and each simulation, $f_\mathrm{obs\_c}$ gives the fraction of successfully completed targets and $f_\mathrm{suc}$ gives the fraction of observed targets that receive their full required exposure, which are thus deemed successful. The ratio of these, $f_\mathrm{obs\_c}/f_\mathrm{suc}$, gives the total fraction of targets observed (not necessarily to completion). The values $f_\mathrm{obs\_c}$ should be compared with the $f_\mathrm{compl}$ value, which represents the fraction of targets that had been expected to reach completion.}
\label{table:targeting_simus}
\centering
\begin{tabular}{l c c c c c c}
\hline\hline
Survey & $f_\mathrm{compl}$ & Base & Raw & Boost & Tilt & Four \\
 & & $f_\mathrm{obs\_c}/f_\mathrm{suc}$ & $f_\mathrm{obs\_c}/f_\mathrm{suc}$ & $f_\mathrm{obs\_c}/f_\mathrm{suc}$ & $f_\mathrm{obs\_c}/f_\mathrm{suc}$ & $f_\mathrm{obs\_c}/f_\mathrm{suc}$ \\
\hline
        LR~1 & 0.6 & 0.72\,/\,0.78 & 0.72\,/\,0.78 & 0.73\,/\,0.79 & 0.72\,/\,0.78 & 0.75\,/\,0.82 \\
        LR~2 & 1.0 & 0.81\,/\,0.96 & 0.81\,/\,0.96 & 0.81\,/\,0.96 & 0.80\,/\,0.96 & 0.81\,/\,0.97 \\
        LR~3 & 0.8 & 0.65\,/\,0.97 & 0.64\,/\,0.97 & 0.65\,/\,0.97 & 0.64\,/\,0.97 & 0.62\,/\,0.97 \\
        LR~4 & 0.9 & 0.91\,/\,0.93 & 0.90\,/\,0.93 & 0.92\,/\,0.94 & 0.91\,/\,0.93 & 0.92\,/\,0.94 \\
        LR~5 & 0.1 & 0.70\,/\,0.74 & 0.79\,/\,0.83 & 0.70\,/\,0.74 & 0.71\,/\,0.75 & 0.65\,/\,0.70 \\
        LR~6 & 0.5 & 0.92\,/\,0.97 & 0.92\,/\,0.96 & 0.92\,/\,0.96 & 0.92\,/\,0.97 & 0.88\,/\,0.94 \\
        \hline
        HR~1 & 0.5 & 0.56\,/\,0.84 & 0.59\,/\,0.85 & 0.56\,/\,0.84 & 0.56\,/\,0.84 & 0.56\,/\,0.82 \\
        HR~2 & 0.5 & 0.39\,/\,0.71 & 0.39\,/\,0.71 & 0.39\,/\,0.71 & 0.39\,/\,0.72 & 0.42\,/\,0.74 \\
        HR~3 & 0.5 & 0.41\,/\,0.59 & 0.38\,/\,0.57 & 0.41\,/\,0.59 & 0.41\,/\,0.60 & 0.41\,/\,0.61 \\
\hline
\end{tabular}
\end{table*}

We selected targets over a fixed sky area and used a fixed tiling pattern as shown in Fig.~\ref{fig:tiles}. The tiling pattern was constructed manually to cover the entire test region with minimal overlap between fields. The tiling pattern in this test is not optimised based on the targets in the test region. Table~\ref{table:surveys} shows the number density and exposure times of each survey. Figure~\ref{fig:example_texp_completeness} also shows how the exposure time of targets vary within a survey. The number density and exposure times vary significantly between surveys, which allowed us to test several aspects of the proposed targeting algorithm. Table~\ref{table:surveys_targeting} shows the mean targeting parameters for each survey. The definition of these parameters is given in Sect.~\ref{sec:fibtotar}.

Figure~\ref{fig:targets} shows the distribution of targets on the sky for the nine surveys used in our test. The target density varies across surveys, but in general it is homogeneous across the chosen area. One exception is LR~3, where targets are highly clustered.
The dense blob (a stellar cluster) that is visible in LR~1 and the slightly higher density region in the middle of HR~1 and HR~2 are features in the Galaxia mock catalogue. We selected these survey regions on purpose to show that surveys affect each other.
Figure~\ref{fig:example_tar_density} shows the target density as a function of smoothing radius. For some surveys, the targets are highly clustered at small scales ($<0.1~$deg), despite the target density being roughly homogeneous across the entire field of view. The clustering at small scales is a challenge for the targeting algorithm since, at small scales, the target density exceeds the fibre density on the sky. This is clearly visible when comparing Fig.~\ref{fig:example_tar_density} with Fig.~\ref{fig:fib_density}.

For test simulations, we used the fixed tiling pattern as shown in Fig.~\ref{fig:tiles}. For the base simulation, we used 15~min exposures six times. Targeting parameters for the base simulation are given in Table~\ref{table:surveys_targeting}. In order to test different aspects of the algorithm we slightly modified the targeting parameters of the base simulation. The description and the deviation from the base simulation is as follows:
\begin{itemize}
    \item \textbf{Base,} the simulation with parameters as described above. Additionally, for each survey $f_\mathrm{add}=1$ and $f_\mathrm{tilt}=1$.
    \item \textbf{Raw} is the same as the base simulation, but $f^\mathrm{sur}_\mathrm{suc}=1$ for each survey. This simulation was used in order to estimate $f^\mathrm{sur}_\mathrm{suc}$ values for each survey.
    \item \textbf{Boost} is the same as the base simulation, but $f_\mathrm{add}=100$ for LR\,2 and LR\,3 and $f_\mathrm{add}=0.01$ for LR\,5 and LR\,6. This simulation was run to test how the additional boost factor affects the completeness of surveys. The factor $f_\mathrm{add}$ was included for surveys that were significantly under- or over-observed in the base simulation.
    \item \textbf{Tilt} is the same as the base simulation, but $f_\mathrm{tilt}(\alpha) = 1-\alpha/\alpha_{\max}$, where $\alpha_{\max}$ is the maximum allowed tilt angle for fibres and $\alpha$ is the tilt angle for a given fibre-target pair.
    \item \textbf{Four} is the same as the base simulation, but it uses 30~min exposures two times plus 15~min exposures two times, instead of 15~min exposures six times.
\end{itemize}
In the next section, we analyse the test simulations and explore various aspects of the targeting algorithm by analysing the simulations that deviated from the base simulation.

\subsection{Results of the test simulations}

Each test simulation is characterised by the fraction of successfully completed targets for each survey. An additional useful parameter to explore is the fraction of successfully completed targets over all observed targets. Both of these values are shown in Table~\ref{table:targeting_simus} for each test simulation and for each survey.

Figure~\ref{fig:example_fibre_efficiency} shows the fibre usage efficiency in the base simulation. Since the number density of targets varies across the selected area (see Fig.~\ref{fig:tiles}), the fibre usage efficiency is not homogeneous. The efficiency is largely determined by the tiling pattern and can be improved if the tiling pattern is estimated using a given set of targets as a prior. In the current paper our aim is to test the fibre to target assignment algorithm for which a fixed tiling pattern is better suited.

\subsubsection{Analysing the base simulation}

We start our analysis with the base simulation. The fraction of successfully completed targets should be compared with the $f_\mathrm{compl}$ value of each survey. We can see that for surveys LR\,2 and LR\,3, the fraction of successfully completed targets is lower than the $f_\mathrm{compl}$ value. Both surveys include highly clustered targets as seen in Table~\ref{table:surveys_targeting}. The mean number of targets per fibre for LR\,2 and LR\,3 is above 25, which is more than three times higher than other surveys. Hence the fraction of successfully completed targets for LR\,2 and LR\,3 is limited by the fixed number density of fibres in the sky.

For surveys LR\,5 and LR\,6, the situation is the opposite; the fraction of successfully completed targets for these surveys is significantly higher than the $f_\mathrm{compl}$ value. For these surveys, the number density of targets is larger than for other surveys, which means that a large fraction of fibres can only be used for targets in these surveys (see Table~\ref{table:surveys_targeting}). Hence there are many fibres that can only be used for these two surveys. This explains why targets from surveys LR\,5 and LR\,6 are targeted more frequently than required as the targeting algorithm attempts to find a target for each fibre.

The target density is very similar across all high-resolution surveys. Nevertheless, the fraction of successfully completed targets is lower than expected for surveys HR\,2 and HR\,3. This can be explained by looking at the exposure times of the high-resolution surveys (see Table~\ref{table:surveys} and Fig.~\ref{fig:example_texp_completeness}). For HR\,1 the exposure times are significantly shorter than for HR\,2 and HR\,3, and targets from HR\,1 are observed more frequently. Since targets from HR\,3 require all six exposures (the target exposure time is close to 90~min), then this limits the number of fibres that can be potentially used for new science targets. Only targets observed during the first exposure can be successfully completed.

In general, the targeting algorithm observes the expected number of targets for each survey. The deviations from the expected number are logically explained and results can only be improved by updating the input catalogues or by improving the tiling pattern. A modification of the targeting algorithm does not significantly improve the results.

\subsubsection{Influence of the factor $f^\mathrm{sur}_\mathrm{suc}$}

As described above, the factor $f^\mathrm{sur}_\mathrm{suc}$ indicates the success of the targeting algorithm for each survey. If the $f^\mathrm{sur}_\mathrm{suc}$ value is lower than 1.0, then the algorithm targets less objects than expected. However if the value is higher than 1.0, then the algorithm targets more objects than expected. In Table~\ref{table:surveys_targeting} we give the $f^\mathrm{sur}_\mathrm{suc}$ values estimated based on the raw simulation setting $f^\mathrm{sur}_\mathrm{suc}=1.0$ for each survey. For the majority of surveys, the value is around 1.0 as expected. The only exception is LR\,5, where the value is significantly higher. This is because the $f_\mathrm{compl}$ value for LR\,5 is only 0.1, and the number density in LR\,5 is relatively high. Hence some of the fibres can only be used for LR\,5 targets.

Comparing the Raw and Base simulations (see Table~\ref{table:targeting_simus}) shows that when including $f^\mathrm{sur}_\mathrm{suc}$, the algorithm reduces the fraction of observed targets for LR\,5. At the same time, the fraction of successfully completed targets is increased for LR\,3 and LR\,4. Hence, the factor $f^\mathrm{sur}_\mathrm{suc}$ balances different surveys. The same effect is visible for high-resolution surveys where some fibres from HR\,1 are moved to HR\,3. In general the inclusion of the factor $f^\mathrm{sur}_\mathrm{suc}$ does what is expected.

\begin{figure}
\centering
\includegraphics[width=88mm]{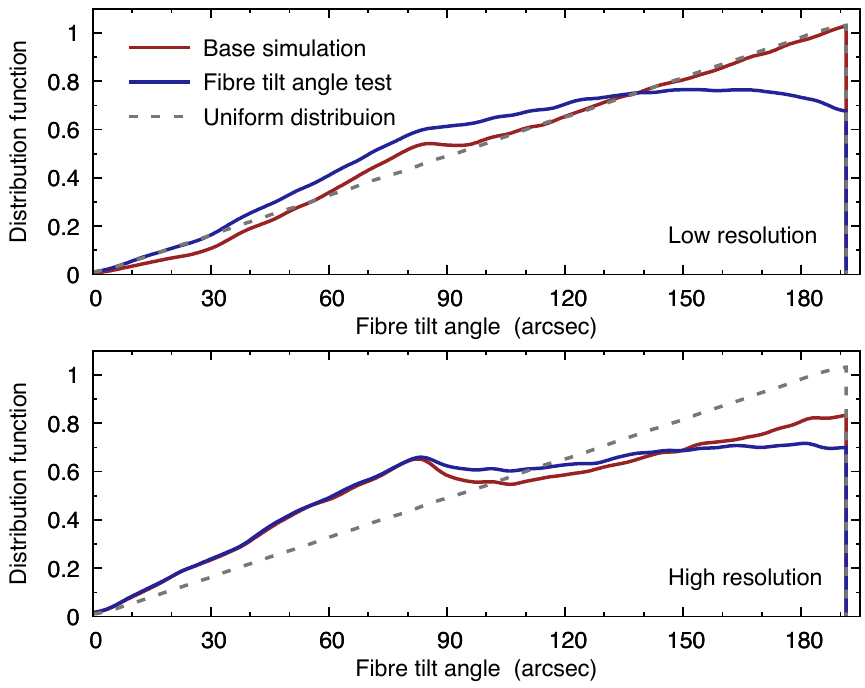}
        \caption{Distribution of tilt angles for low- (upper panel) and high-resolution (lower panel) spectrograph fibres. Tilt angle was measured from the fibre home position. The red line shows the distribution for the base simulation whilst the blue line shows the tilt angle distribution after introducing the non-constant $f_\mathrm{tilt}(\alpha)$ function (see Sect.~\ref{sec:influence_tilt} for more details). The dashed line shows the uniform distribution of tilt angles. The local bump around 80~arcsec is due to the fixed fibre pattern (see Fig.~\ref{fig:fibres_multi}).}
        \label{fig:example_tilt_distribution}
\end{figure}

\begin{figure}
\centering
\includegraphics[width=88mm]{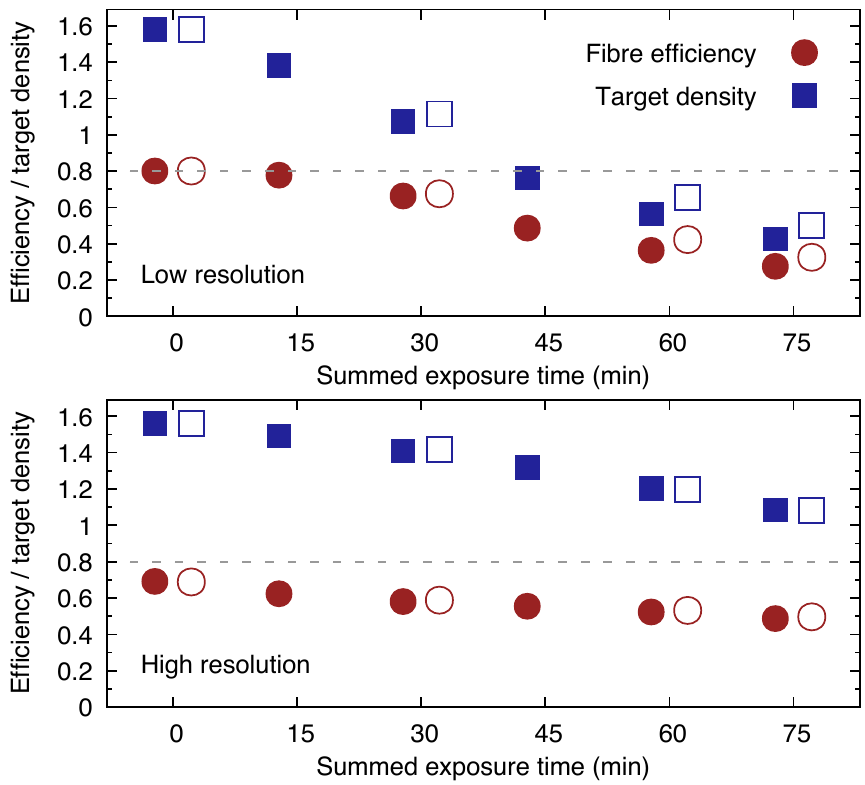}
        \caption{Fibre usage efficiency (red circles) for low- (upper panel) and high-resolution (lower panel) surveys as a function of survey progress. For both resolutions, we used only 80 percent of fibres for science targets, where the rest of the fibres were reserved for standard stars and the sky background. Hence the maximum fibre efficiency is 0.8. Blue squares show the mean target density on the sky normalised by the number density of science fibres. At the beginning of the survey, the target density exceeds the fibre density by two times, but for the last exposures the target density in the low-resolution regime is lower (less than 1.0) than the fibre density. Hence, for the last exposures we do not have enough targets to fill all fibres. Filled points represent the base simulation (15~min exposure six times) and empty points indicate the simulation with four visits (30~min twice plus 15~min twice).}
        \label{fig:example_fibre_efficiency_progress}
\end{figure}

\subsubsection{Influence of the factor $f_\mathrm{add}$}

By analysing the base simulation, we can see that for surveys LR~2 and LR~3 the completion fraction is lower than expected; while for surveys LR~5 and LR~6, it is significantly higher than requested by $f_\mathrm{compl}$. To determine how the performance of these surveys can be improved, we used $f_\mathrm{add}=100$ for LR~2 and LR~3, and we set $f_\mathrm{add}=0.01$ for LR~5 and LR~6.

Table~\ref{table:targeting_simus} shows that including the additional boost factor for surveys LR~2 and LR~3 does not increase the completeness of these surveys. This means that the completeness of these surveys is primarily limited by the fixed number density of fibres and the influence of other surveys is marginal. The completeness of LR~2 and LR~3 surveys can only be increased when increasing the number of exposures, which increases the total number of fibres per sky area.

By analysing surveys LR~5 and LR~6 where the boost factor was set to $f_\mathrm{add}=0.01$, we see that the completeness is only marginally affected (i.e. less than 1~\%). This can be explained by looking at Table~\ref{table:surveys_targeting}. The number of accessible fibres for LR~6 is the highest, while for most of the other surveys it is significantly lower. This means that there are fibres that can only reach LR~6 targets and cannot be used for other surveys. Since fibre usage is maximised in the targeting algorithm, the targets from LR~6 are more often observed than requested. The situation is similar for LR5, but, additionally, the requested completion fraction for LR~5 is very low (0.1), which also plays a key role.

Adding a boost factor for some surveys has a side effect. Since some fibres are taken away from surveys LR~5 and LR~6, the completion fraction for other surveys (LR~1 and LR~4) is slightly increased. Whether this is a positive or negative effect depends on the viewpoint of the survey science case. In general, the $f_\mathrm{add}$ factor can be used to fine tune the balance between surveys if necessary.

\begin{figure*}
\centering
\includegraphics[width=180mm]{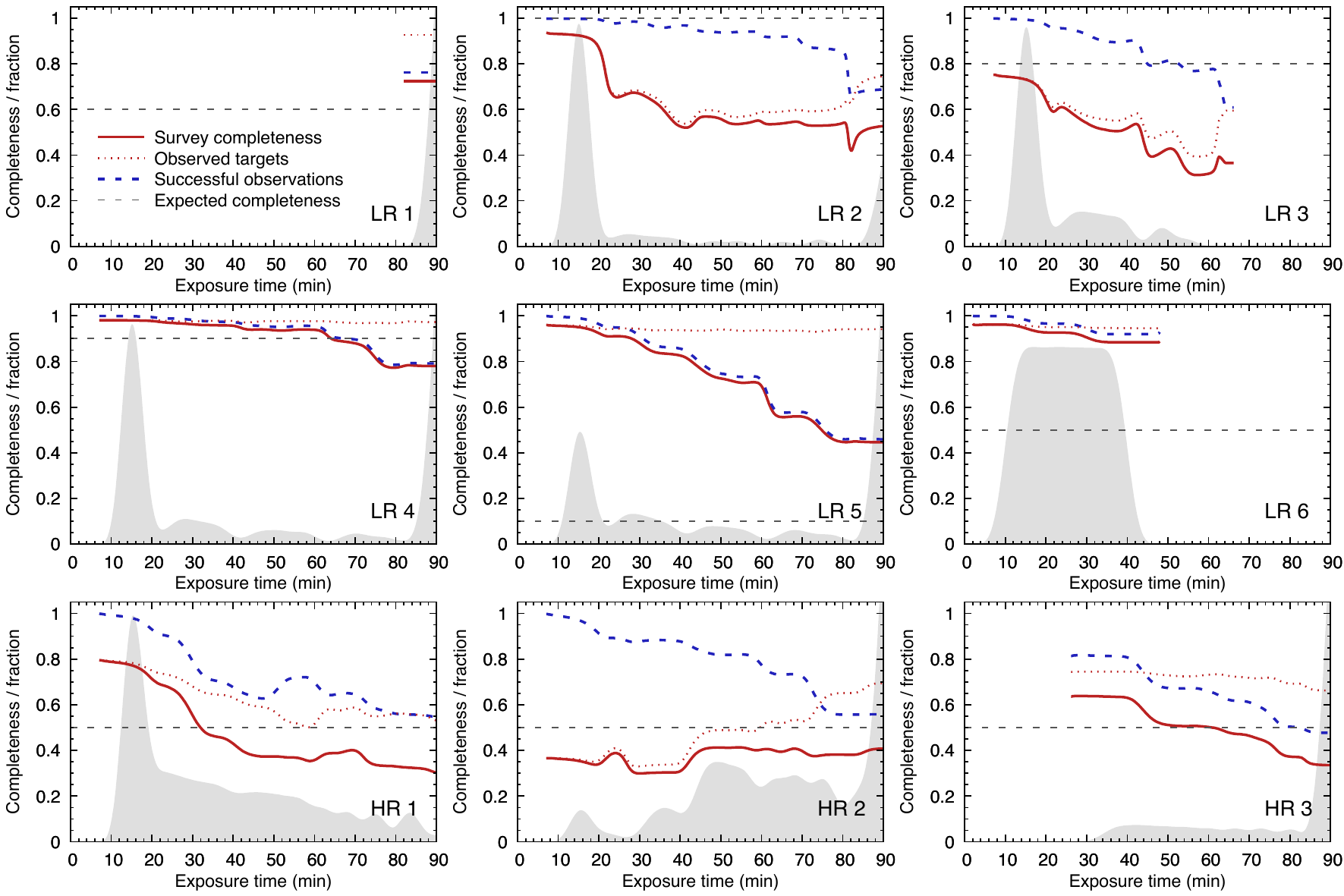}
        \caption{Survey completeness as a function of target exposure time. The grey area shows the distribution of exposure times for each survey. The dashed horizontal line shows the required completeness fraction ($f_\mathrm{compl}$ value) for each survey. The dotted red line shows the fraction of observed targets for each survey whilst the solid red line shows the fraction of successfully completed targets. The dashed blue line shows the fraction of successfully completed targets over all observed targets.}
        \label{fig:example_texp_completeness}
\end{figure*}

\begin{figure*}
\centering
\includegraphics[width=180mm]{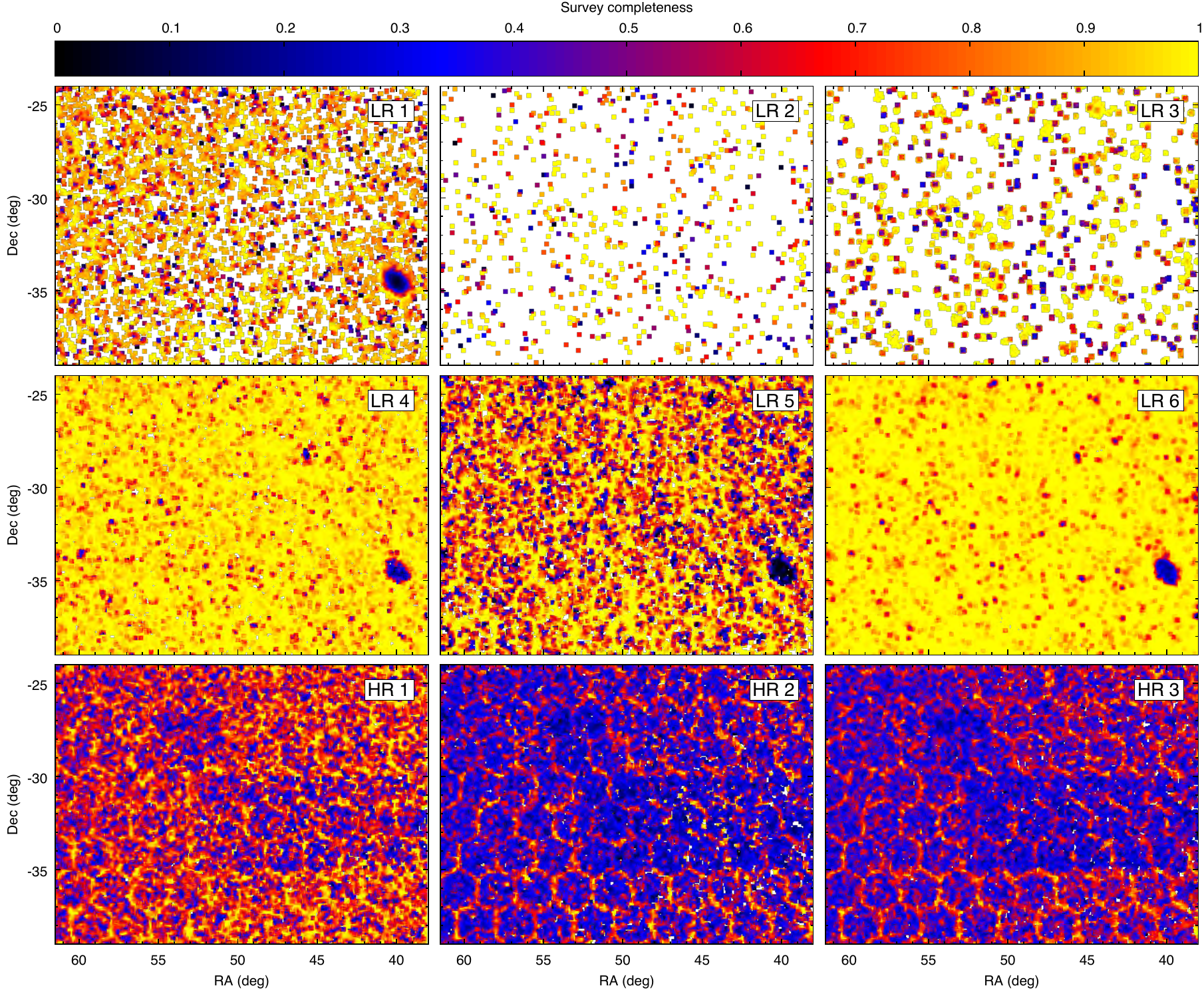}
        \caption{Fraction of successfully completed targets out of all targets as a function of sky coordinates for each survey. The tiling pattern is clearly visible in the selection function. Additionally if there is a feature (a denser region) in one survey, it leaves a footprint in all other surveys. However, the feature visible in one survey mainly affects surveys with the same resolution. Hence, the cluster visible in LR surveys (lower right corner) does not affect the completeness for HR surveys.}
        \label{fig:example_map_completeness}
\end{figure*}

\begin{figure*}
\centering
\includegraphics[width=180mm]{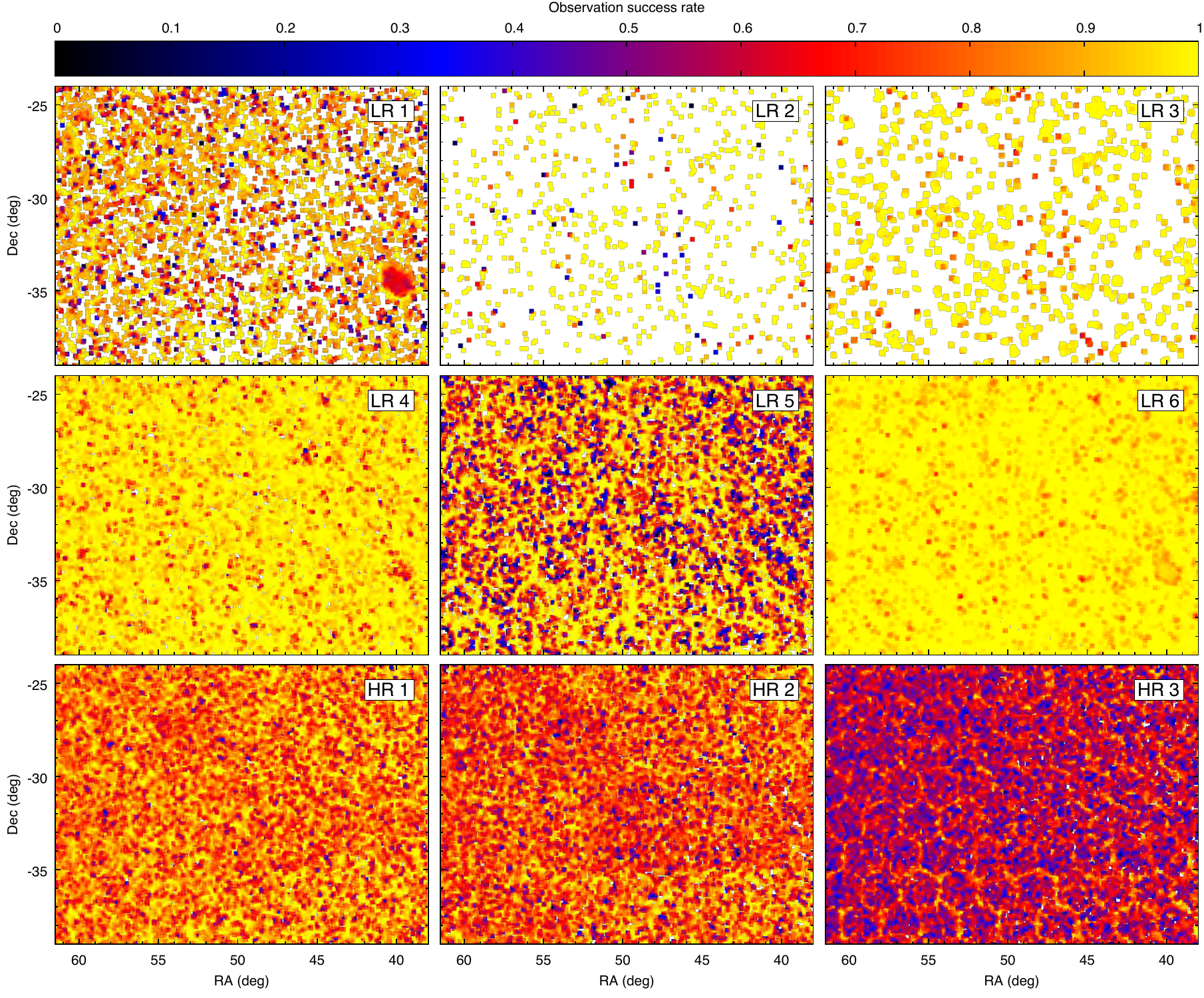}
        \caption{Fraction of successfully observed targets over all observed targets. Yellow points show successfully completed targets. The tiling pattern is clearly visible for surveys that contain a large number of targets with long requested exposure times (e.g. HR~3 in the bottom right panel). For surveys with targets requiring short exposures, the tiling pattern is not as pronounced.}
        \label{fig:example_map_success}
\end{figure*}

\begin{figure}
\centering
\includegraphics[width=88mm]{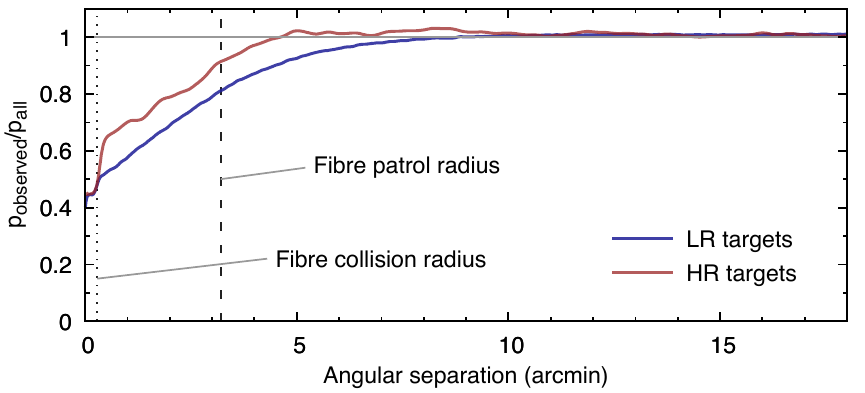}
        \caption{Selection of targets as a function of angular separation. The distribution of angular separation of observed targets was divided by the distribution of angular separation of all targets. The blue line shows low-resolution targets and the red line shows high-resolution targets. The vertical dashed line shows the fibre patrol radius (3.2~arcmin) and the vertical dotted line shows the fibre collision distance (17~arcsec). Due to the fixed fibre pattern, targets with smaller angular separation are less likely to be targeted. Targeting efficiency rapidly drops for targets closer than fibre collision distance.}
        \label{fig:angular_separation}
\end{figure}

\subsubsection{Influence of the factor $f_\mathrm{tilt}$}
\label{sec:influence_tilt}

Due to the fibre positioner design (Echidna fibre-positioning technology), fibre throughput is reduced if fibres are tilted from their home position. The instrument is most effective if targets are allocated close to the fibre home positions. In the proposed targeting algorithm, targets that are close to the fibre home position can be preferred by using a weight factor $f_\mathrm{tilt}$. The results where targets close to fibre home positions were preferentially selected are shown in Table~\ref{table:targeting_simus}, in the column Tilt. For survey completion, the effect of $f_\mathrm{tilt}$ is less than 1~\% for all surveys.

Figure~\ref{fig:example_tilt_distribution} shows the fibre tilt angle distribution for the base simulation and for a simulation that prefers targets close to the fibre home position. We can see that introducing a factor $f_\mathrm{tilt}$ has a significant effect on low-resolution spectrograph fibres. The effect is much smaller for high-resolution spectrograph fibres because their flexibility is lower. For high-resolution spectrograph fibres roughly 40~\% of the field of view can only be reached by a single fibre (see Table~\ref{table:fib_coverage}).

Although the introduction of $f_\mathrm{tilt}$ prefers targets that are close to the fibre home positions, it should still be evaluated as to whether and how this affects survey scientific goals. Introducing the factor $f_\mathrm{tilt}$ affects the selection function at small scales.

\subsubsection{Influence of the number of visits}

For an optimal tiling pattern, we should combine exposures with different lengths. It is clear that in some sky regions the required exposure time is determined by the objects with the longest exposures, whilst in other sky regions the required exposure time is determined by a high density of targets with low exposure times.

For the base simulation, we used 15~min exposures six times. To test the combination of different exposure times, we ran the algorithm using 30~min exposures two times and 15~min exposures two times. The total exposure time (90~min) for both cases is the same. The mean target density before each visit, which is normalised by the fibre density, is shown in Fig.~\ref{fig:example_fibre_efficiency_progress}. For high-resolution, the target density is always higher than fibre density; whilst for low-resolution, the target density for later visits is lower than the fibre density. This is due to most of the high-resolution targets requiring long exposures and the number density of remaining targets not dropping rapidly. In the low-resolution regime the majority of targets require short exposures; for low-resolution, the number density of remaining targets does drop rapidly.

Figure~\ref{fig:example_fibre_efficiency_progress} shows that the fibre usage efficiency is almost independent of the number of visits, that is, six or four. Table~\ref{table:targeting_simus} shows that the survey completion fraction only slightly depends on the number of visits. For some surveys the completion fraction is slightly better with four visits whilst for other surveys it is the opposite. In general there are no large differences between these two cases. From a practical standpoint, a smaller number of visits is preferred as there is an additional overhead associated with each exposure.

\subsection{Analysing selection functions}
\label{sec:selfunc}

The selection function defines the survey completeness as a function of some parameter, such as sky coordinates or magnitude. To understand the targeting algorithm, we analysed the selection function as a function of target exposure time and as a function of sky coordinates. In our test simulations, the expected completeness as a function of these parameters should be uniform.

Figure~\ref{fig:example_texp_completeness} shows the completeness of each survey as a function of target exposure time. The horizontal dashed line shows the expected completeness, and the solid red line shows the fraction of successfully completed targets. Ideally this should match the dashed horizontal line. The completeness for short exposure targets is higher than for long exposure targets. This is expected as during later visits, the empty fibres are used to observe short exposure targets. For long exposure targets, the completeness is roughly constant, as required, and the targeting algorithm tries to achieve uniform completeness regardless of the target exposure time, or magnitude. This cannot be achieved perfectly, mainly due to the fixed fibre density and clustering of targets; we note that for some fibres there are more targets than for other fibres. In general, the achieved uniformity with regards to exposure time is not perfect, but it is nearly independent of the exposure time distribution (grey shaded areas on Fig.~\ref{fig:example_texp_completeness}). The remaining dependence on exposure time is mainly caused by the fixed fibre pattern of the 4MOST instrument, by the distribution of targets and their exposure times, and by the used tiling pattern. Full survey strategy optimisation is required to balance the survey input catalogues and to find the optimal tiling solution.

By comparing the expected completeness (horizontal dashed line) with survey completeness (solid red line), we see that the survey completeness of LR~2 and LR~3 is lower than expected. Both surveys are galaxy cluster surveys and the number of successfully completed targets is limited by the number density of fibres. However, everything that is possible to observe with six visits is observed. For high resolution, the completeness is slightly lower than expected. This is because most of the high-resolution targets require long exposures and with one fibre you cannot observe two targets that require more than a 45~min exposure each. Taking that into account, the completion fraction for high-resolution targets is also limited by the fibre density. The targeting algorithm balances different surveys in a way that each survey is under-observed in a roughly equal manner. The targeting algorithm does not prefer one survey to another.

The question is inevitably posed as to how we can improve the completion of surveys. For surveys that require short exposures, we can increase the number of visits whilst reducing the exposure time per visit. Even if the total exposure time is the same, we achieve better completeness as there is more freedom to place fibres on targets. However, with short exposures we increase the noise due to the larger numbers of read-outs and we have to observe for longer periods of time to achieve the same signal-to-noise ratio. Hence, the optimal solution is not straightforward. For surveys that require long exposures, the only way to improve the completion is to increase the total exposure time. In the current setup, if most of the targets require more than 45~min exposures, then it does not matter whether they are observed six times for 15~min, three times for 30~min, or two times for 45~min; the completion is roughly the same in all three cases. Determining the optimal number of visits and exposure times is a complicated optimisation problem, which is not further discussed in this paper.

Figure~\ref{fig:example_map_completeness} shows the fraction of successfully observed targets as a function of sky coordinates for each survey. At scales larger than the field of view, the completeness is homogeneous, which is  as expected. For surveys where the completeness is less than 1.0, we can see a visible tiling pattern in the sky. In regions where tiles overlap, the completeness is significantly higher. Additionally we see that if there is a high density region in one survey, such that there is  a dense blob of targets to the right of the field, for example, then the completeness is affected in all surveys. The targeting algorithm tries to observe all targets equally, and if tiling does not allow for this, then all surveys are penalised equally. This emphasises that for a homogeneous selection function, the tiling pattern should match the target density. In looking at the target distribution for high-resolution surveys (see Fig.~\ref{fig:targets}), we can see that for HR~1 and HR~2 there is a clear pattern present in the middle of the field. Although this pattern is not present in HR~3, it is clearly visible in the completeness map (see Fig.~\ref{fig:example_map_completeness}). In principle, the footprint of all surveys and all inhomogeneities are visible in the final completeness map of all surveys. This is a side effect of all surveys sharing the focal plane and observing targets simultaneously. The increase in survey efficiency comes with the price that the selection function is more complicated.

Figure~\ref{fig:example_map_success} shows the fraction of successfully observed targets over all observed targets. From this map, we see that most of the targets that are observed are successfully completed. This success rate is homogeneous and it barely depends on the target density. For example, the dense cloud of low-resolution targets does not affect the success rate. The tiling footprint is most strongly visible for HR~3 and LR~5. These are the surveys where most of the targets require long exposures. For all other surveys the success rate depends only very weakly on the tiling pattern. This is expected as the targeting algorithm tries to only observe targets that can be successfully completed. The targeting algorithm only observes targets that cannot be successfully completed if no other options are available.

Figure~\ref{fig:angular_separation} shows the targeting efficiency as a function of angular separation between targets. In an ideal case, the selection of targets should be independent of the angular separation of targets. Figure~\ref{fig:angular_separation} shows that targeting efficiency decreases for targets where angular separation is less than $\sim 5$~arcmin. This is the outcome of a fixed fibre pattern and a limited patrol area around each fibre. There is an additional decrease of efficiency at very small scales (17~arcsec), which is caused by the fibre collision issue; two fibres in a single field of view cannot be placed closer than 17~arcsec. Figure~\ref{fig:angular_separation} clearly shows that the fixed fibre pattern has a more severe effect than the fibre collision issue. In a forthcoming paper, we will analyse how these effects affect the three-dimensional clustering measurements. We will use the methodology presented in \citet{2017MNRAS.472.1106B} and analyse how well we can recover the galaxy two-point correlation function in a 4MOST cosmology redshift survey.

\section{Conclusion and discussion}
\label{sec:conclusion}

In this paper we propose a probabilistic fibre-to-target assignment algorithm that can deal with concurrent surveys that are observed simultaneously. The proposed algorithm takes the limitations of a fibre positioner system into account (Echidna in case of 4MOST), and it can deal with surveys with different targeting requirements and number densities. One aim of the proposed algorithm is to maximise the target completion, thus minimising the number of half-completed targets, while achieving survey completion that is independent of the target exposure time or magnitude. Due to the fixed fibre pattern of the 4MOST instrument, the targeting completeness at small scales is not random. In fact, the fixed fibre pattern of the 4MOST instrument has a much larger impact than the fibre collision issue, which only has an effect at very small scales. The probabilistic approach minimises a fixed fibre pattern effect and allows one to achieve nearly uniform targeting completeness even at small angular scales. An additional benefit of the proposed algorithm is that the targeting completeness with respect to the target exposure time is homogeneous, hence, the algorithm does not penalise long-exposure targets nor prefer short-exposure targets.

In general, the free parameters and functions, such as  $f_\mathrm{tilt}$, that are introduced in the algorithm can be used to control certain aspects of the survey optimisation problem. The optimal choice of these parameters and functions should be determined during the survey preparation and optimisation. The need and value of these parameters depend on the instrument characteristics and input target catalogues.

Computationally, the proposed algorithm works reasonably fast. For example, when taking the full five-year 4MOST survey that contains about 40 million targets and 40\,000 individual observations on a decent server with 18 cores and 256~GB of memory, it takes about 20 minutes to calculate the fibre-target pair probabilities. Then it takes less than 5 minutes to simulate the five-year survey observations. Hence, it is reasonable to use the computational cost during the real 4MOST observations.

The proposed algorithm assumes that the tiling pattern is fixed and known beforehand. The generation of an optimal tiling pattern is a critical step and ultimately decides the summed exposure time and number of visits for a given sky region. Finding an optimal tiling pattern for a given set of targets is a complicated optimisation problem \citep[see for example][]{2003AJ....125.2276B}. \citet{2010PASA...27...76R} developed the Greedy algorithm that works efficiently for spatially dense surveys. Due to the low number density of targets in some sky areas, the Greedy algorithm does not work everywhere in the 4MOST footprint. We will address the optimal tiling pattern question in a future paper, where we model the tiling pattern as a marked point process. In this framework, the tiling pattern is seen as a configuration of random interacting objects driven by the probability density of a marked point process. The solution of the optimal tiling problem is given by the construction and manipulation of such a probability density. Details of this algorithm are given in Tempel et al. (in prep). 

The current paper addresses the key aspects of the fibre-to-target assignment problem. However, there are several other aspects that should be considered during real observations. All of these are potentially useful improvements that would extend beyond the proposed algorithm. Below we mention and discuss some of these aspects.
\paragraph{\textbf{Realistic target progress updates.}} In this paper target progress is estimated based upon the required and observed exposure times. In reality, targets are observed during several sky conditions, and the summed exposure time may not be sufficient to track target progress. In addition to monitoring the exposure time per target, 4MOST also monitors the signal-to-noise ratio of the observed spectrum or redshift success as a criterion of target progress. These aspects are important and should be included in the algorithm, depending on what is desired for quantifying success during the observations.
\paragraph{\textbf{Duplicated targets.}} During real observations, one target can be present in many surveys. Depending on the survey's scientific goals, exposure time requirements for the same target can be different for different surveys. A straightforward way to take this into account in the targeting algorithm is to include the target in all survey input catalogues. This allows us to calculate correct probabilities for fibre-target pairs. To avoid observing the same target multiple times, the target should be removed from all input catalogues once the target is successfully completed. The success of the target should be estimated based on the most demanding survey requirement.
\paragraph{\textbf{Prioritising specific targets.}} In the proposed algorithm, targets are not specifically prioritised. However, there might be a scientifically justified need to prioritise certain targets in some surveys. One example is the clusters survey in which a cluster main galaxy is scientifically more important than the cluster member galaxies. Whether this can be automatically solved by the probabilistic targeting algorithm or whether it needs special tuning requires additional analysis. This also depends on the specific requirements of the survey.
\paragraph{\textbf{Transients in the input catalogue.}} The calculation of fibre-target pair priorities assumes that the list of all targets is known and does not change during observations. However, this assumption is violated if we want to observe transients. The list of targets is not known beforehand and cannot be included in the probability calculations. We expect that if the fraction of transients is small, then this has a negligible effect on the main surveys. However, this should be tested using real mock catalogues. 
\paragraph{\textbf{Repeated observations of the same target.}} In the proposed algorithm, repeated observations of targets are prioritised to guarantee that they are successfully completed. Since exposure times and sky conditions of single observations are different, some observations are more efficient than others to successfully complete partially observed targets. To increase survey efficiency, this should be taken into account in the proposed algorithm.
\paragraph{\textbf{Cross-talk.}} If faint and bright object are observed side-by-side (neighbouring traces on the CCD), then the light from the bright object affects the spectrum of the faint object. If cross-talk is a serious issue, then selecting bright and faint objects that are neighbours on the CCD should be avoided with the targeting algorithm.
\paragraph{\textbf{Optimising fibre usage efficiency.}} In the current paper, the fibre allocation starts with fibres with the lowest number of potential targets. This is a very simple implementation and tends to maximise the fibre usage. However the fibre usage efficiency can be increased by using a Monte-Carlo method with simulated annealing during the fibre-to-target assignment. This increases the computational cost of the algorithm and requires further analysis to test whether increased computational cost provides a sufficient increase in efficiency to be justified.
\paragraph{\textbf{Field order and cadence.}} Fibre-target pair probability calculation assumes that field order is known. In many cases this is not known beforehand and it is determined during real observations depending, for example, on the sky conditions. To overcome this problem, the targeting algorithm can use a random field order, or a best estimate. This has some effect on the survey efficiency. A cadence, which is the required minimum time difference between repeated observations or the requirement that a target should be completed during one night, is another factor that affects the field order. However, it does not have a direct impact on the fibre-to-target assignment algorithm.

\paragraph{}
To conclude, the algorithm presented in this paper is only a baseline and it does not yet address all aspects of real-life observations that should be considered. Most of these can be tested using more realistic target catalogues. In preparation for the 4MOST survey, we will test these effects and modify the targeting algorithm as needed. The current algorithm is a proposed solution for the 4MOST survey and based on tests presented in this paper, it provides a promising solution for multi-fibre spectroscopic surveys with several concurrent sub-surveys. With the appropriate modifications, the proposed approach can be applied to any other multi-object survey. The required modifications in a fibre-to-target probability assignment for other instruments and surveys depend on the instrument's capabilities and survey science cases.

\begin{acknowledgements}
This work has made use of the development effort for 4MOST, an instrument under construction by the 4MOST Consortium (\url{https://www.4most.eu/cms/consortium/}) for the European Southern Observatory (ESO). Part of this work was supported by institutional research funding IUT26-2 and IUT40-2 of the Estonian Ministry of Education and Research. We acknowledge the support by the Centre of Excellence ``Dark side of the Universe'' (TK133) and by the grant MOBTP86 financed by the European Union through the European Regional Development Fund. P.N. acknowledges support from the STFC through ST/P000541/1 and the Royal Society through the award of a University Research Fellowship. M.-R.L.C. acknowledges funding from the European Research Council (ERC) under the European Union's Horizon 2020 research and innovation programme (grant agreement nr. 682115). A.K. acknowledges support from the Sonderforschungsbereich SFB 881 ``The Milky Way System'' (subprojects A03, A05, A08) of the DFG. M.K. acknowledges support from DLR grant 50OR1904. B.F.R acknowledges support by MNiSW grant DIR/WK/2018/12 and by grant 314 at the PSNC. G.T. was supported by the grant The New Milky Way from the Knut and Alice Wallenberg foundation, by the grant 2016-03412 from the Swedish Research Council, and by the Swedish strategic research programme eSSENCE.
\end{acknowledgements}

\bibliographystyle{aa_2019_ArXiv} 
\bibliography{mybib} 

\begin{thebibliography}{29}
\expandafter\ifx\csname natexlab\endcsname\relax\def\natexlab#1{#1}\fi

\bibitem[{{Bensby} {et~al.}(2019){Bensby}, {Bergemann}, {Rybizki}, {Lemasle},
  {Howes}, {Kovalev}, {Agertz}, {Asplund}, {Barklem}, {Battistini},
  {Casagrande}, {Chiappini}, {Church}, {Feltzing}, {Ford}, {Gerhard},
  {Kushniruk}, {Kordopatis}, {Lind}, {Minchev}, {McMillan}, {Rix}, {Ryde}, \&
  {Traven}}]{2019Msngr.175...35B}
{Bensby}, T., {Bergemann}, M., {Rybizki}, J., {et~al.} 2019, The Messenger,
  175, 35, \eprint{1903.02470}

\bibitem[{{Bianchi} {et~al.}(2018){Bianchi}, {Burden}, {Percival}, {Brooks},
  {Cahn}, {Forero-Romero}, {Levi}, {Ross}, \& {Tarle}}]{2018MNRAS.481.2338B}
{Bianchi}, D., {Burden}, A., {Percival}, W.~J., {et~al.} 2018, \mnras, 481,
  2338, \eprint{1805.00951}

\bibitem[{{Bianchi} \& {Percival}(2017)}]{2017MNRAS.472.1106B}
{Bianchi}, D. \& {Percival}, W.~J. 2017, \mnras, 472, 1106, \eprint{1703.02070}

\bibitem[{{Blanton} {et~al.}(2003){Blanton}, {Lin}, {Lupton}, {Maley}, {Young},
  {Zehavi}, \& {Loveday}}]{2003AJ....125.2276B}
{Blanton}, M.~R., {Lin}, H., {Lupton}, R.~H., {et~al.} 2003, \aj, 125, 2276,
  \eprint{astro-ph/0105535}

\bibitem[{{Brzeski} {et~al.}(2018){Brzeski}, {Baker}, {Baker}, {Brown}, {Case},
  {Farrell}, {Gillingham}, {Klauser}, {Lawrence}, {Mali}, {Muller}, {Nichani},
  {Pai}, {Smedley}, {Venkatesan}, \& {Waller}}]{2018SPIE10702E..79B}
{Brzeski}, J., {Baker}, G., {Baker}, S., {et~al.} 2018, in Society of
  Photo-Optical Instrumentation Engineers (SPIE) Conference Series, Vol. 10702,
  \procspie, 1070279

\bibitem[{{Chiappini} {et~al.}(2019){Chiappini}, {Minchev}, {Starkenburg},
  {Anders}, {Fusillo}, {Gerhard}, {Guiglion}, {Khalatyan}, {Kordopatis},
  {Lemasle}, {Matijevic}, {Queiroz}, {Schwope}, {Steinmetz}, {Storm}, {Traven},
  {Tremblay}, {Valentini}, {Andrae}, {Arentsen}, {Asplund}, {Bensby},
  {Bergemann}, {Casagrande}, {Church}, {Cescutti}, {Feltzing}, {Fouesneau},
  {Grebel}, {Kovalev}, {McMillan}, {Monari}, {Rybizki}, {Ryde}, {Rix},
  {Walton}, {Xiang}, {Zucker}, \& {4MIDABLE-Lr Team}}]{2019Msngr.175...30C}
{Chiappini}, C., {Minchev}, I., {Starkenburg}, E., {et~al.} 2019, The
  Messenger, 175, 30, \eprint{1903.02469}

\bibitem[{{Christlieb} {et~al.}(2019){Christlieb}, {Battistini}, {Bonifacio},
  {Caffau}, {Ludwig}, {Asplund}, {Barklem}, {Bergemann}, {Church}, {Feltzing},
  {Ford}, {Grebel}, {Hansen}, {Helmi}, {Kordopatis}, {Kovalev}, {Korn}, {Lind},
  {Quirrenbach}, {Rybizki}, {Sk{\'u}lad{\'o}ttir}, \&
  {Starkenburg}}]{2019Msngr.175...26C}
{Christlieb}, N., {Battistini}, C., {Bonifacio}, P., {et~al.} 2019, The
  Messenger, 175, 26, \eprint{1903.02468}

\bibitem[{{Cioni} {et~al.}(2019){Cioni}, {Storm}, {Bell}, {Lemasle},
  {Niederhofer}, {Bestenlehner}, {El Youssoufi}, {Feltzing},
  {Gonz{\'a}lez-Fern{\'a}ndez}, {Grebel}, {Hobbs}, {Irwin}, {Jablonka}, {Koch},
  {Schnurr}, {Schmidt}, \& {Steinmetz}}]{2019Msngr.175...54C}
{Cioni}, M.-.~R.~L., {Storm}, J., {Bell}, C.~P.~M., {et~al.} 2019, The
  Messenger, 175, 54, \eprint{1903.02475}

\bibitem[{{Cirasuolo} \& {MOONS Consortium}(2016)}]{2016ASPC..507..109C}
{Cirasuolo}, M. \& {MOONS Consortium}. 2016, in Astronomical Society of the
  Pacific Conference Series, Vol. 507, Multi-Object Spectroscopy in the Next
  Decade: Big Questions, Large Surveys, and Wide Fields, ed. I.~{Skillen},
  M.~{Balcells}, \& S.~{Trager}, 109

\bibitem[{{Comparat} {et~al.}(2019){Comparat}, {Merloni}, {Salvato}, {Nandra},
  {Boller}, {Georgakakis}, {Finoguenov}, {Dwelly}, {Buchner}, {Del Moro},
  {Clerc}, {Wang}, {Zhao}, {Prada}, {Yepes}, {Brusa}, {Krumpe}, \&
  {Liu}}]{2019MNRAS.tmp.1335C}
{Comparat}, J., {Merloni}, A., {Salvato}, M., {et~al.} 2019, \mnras, 1335,
  \eprint{1901.10866}

\bibitem[{{Dalton} {et~al.}(2012){Dalton}, {Trager}, {Abrams}, {Carter},
  {Bonifacio}, {Aguerri}, {MacIntosh}, {Evans}, {Lewis}, {Navarro}, {Agocs},
  {Dee}, {Rousset}, {Tosh}, {Middleton}, {Pragt}, {Terrett}, {Brock}, {Benn},
  {Verheijen}, {Cano Infantes}, {Bevil}, {Steele}, {Mottram}, {Bates},
  {Gribbin}, {Rey}, {Rodriguez}, {Delgado}, {Guinouard}, {Walton}, {Irwin},
  {Jagourel}, {Stuik}, {Gerlofsma}, {Roelfsma}, {Skillen}, {Ridings},
  {Balcells}, {Daban}, {Gouvret}, {Venema}, \& {Girard}}]{2012SPIE.8446E..0PD}
{Dalton}, G., {Trager}, S.~C., {Abrams}, D.~C., {et~al.} 2012, in Society of
  Photo-Optical Instrumentation Engineers (SPIE) Conference Series, Vol. 8446,
  Ground-based and Airborne Instrumentation for Astronomy IV, 84460P

\bibitem[{{de Jong} {et~al.}(2019){de Jong}, {Agertz}, {Berbel}, {Aird},
  {Alexander}, {Amarsi}, {Anders}, {Andrae}, {Ansarinejad}, {Ansorge},
  {Antilogus}, {Anwand -Heerwart}, {Arentsen}, {Arnadottir}, {Asplund},
  {Auger}, {Azais}, {Baade}, {Baker}, {Baker}, {Balbinot}, {Baldry}, {Banerji},
  {Barden}, {Barklem}, {Barth{\'e}l{\'e}my-Mazot}, {Battistini}, {Bauer},
  {Bell}, {Bellido-Tirado}, {Bellstedt}, {Belokurov}, {Bensby}, {Bergemann},
  {Bestenlehner}, {Bielby}, {Bilicki}, {Blake}, {Bland-Hawthorn}, {Boeche},
  {Boland}, {Boller}, {Bongard}, {Bongiorno}, {Bonifacio}, {Boudon}, {Brooks},
  {Brown}, {Brown}, {Br{\"u}ggen}, {Brynnel}, {Brzeski}, {Buchert},
  {Buschkamp}, {Caffau}, {Caillier}, {Carrick}, {Casagrande}, {Case}, {Casey},
  {Cesarini}, {Cescutti}, {Chapuis}, {Chiappini}, {Childress}, {Christlieb},
  {Church}, {Cioni}, {Cluver}, {Colless}, {Collett}, {Comparat}, {Cooper},
  {Couch}, {Courbin}, {Croom}, {Croton}, {Daguis{\'e}}, {Dalton}, {Davies},
  {Davis}, {de Laverny}, {Deason}, {Dionies}, {Disseau}, {Doel}, {D{\"o}scher},
  {Driver}, {Dwelly}, {Eckert}, {Edge}, {Edvardsson}, {Youssoufi}, {Elhaddad},
  {Enke}, {Erfanianfar}, {Farrell}, {Fechner}, {Feiz}, {Feltzing}, {Ferreras},
  {Feuerstein}, {Feuillet}, {Finoguenov}, {Ford}, {Fotopoulou}, {Fouesneau},
  {Frenk}, {Frey}, {Gaessler}, {Geier}, {Fusillo}, {Gerhard}, {Giannantonio},
  {Giannone}, {Gibson}, {Gillingham}, {Gonz{\'a}lez-Fern{\'a}ndez},
  {Gonzalez-Solares}, {Gottloeber}, {Gould}, {Grebel}, {Gueguen}, {Guiglion},
  {Haehnelt}, {Hahn}, {Hansen}, {Hartman}, {Hauptner}, {Hawkins}, {Haynes},
  {Haynes}, {Heiter}, {Helmi}, {Aguayo}, {Hewett}, {Hinton}, {Hobbs}, {Hoenig},
  {Hofman}, {Hook}, {Hopgood}, {Hopkins}, {Hourihane}, {Howes}, {Howlett},
  {Huet}, {Irwin}, {Iwert}, {Jablonka}, {Jahn}, {Jahnke}, {Jarno}, {Jin},
  {Jofre}, {Johl}, {Jones}, {J{\"o}nsson}, {Jordan}, {Karovicova}, {Khalatyan},
  {Kelz}, {Kennicutt}, {King}, {Kitaura}, {Klar}, {Klauser}, {Kneib}, {Koch},
  {Koposov}, {Kordopatis}, {Korn}, {Kosmalski}, {Kotak}, {Kovalev}, {Kreckel},
  {Kripak}, {Krumpe}, {Kuijken}, {Kunder}, {Kushniruk}, {Lam}, {Lamer},
  {Laurent}, {Lawrence}, {Lehmitz}, {Lemasle}, {Lewis}, {Li}, {Lidman}, {Lind},
  {Liske}, {Lizon}, {Loveday}, {Ludwig}, {McDermid}, {Maguire}, {Mainieri},
  {Mali}, {Mandel}, {Mandel}, {Mannering}, {Martell}, {Martinez Delgado},
  {Matijevic}, {McGregor}, {McMahon}, {McMillan}, {Mena}, {Merloni}, {Meyer},
  {Michel}, {Micheva}, {Migniau}, {Minchev}, {Monari}, {Muller}, {Murphy},
  {Muthukrishna}, {Nandra}, {Navarro}, {Ness}, {Nichani}, {Nichol}, {Nicklas},
  {Niederhofer}, {Norberg}, {Obreschkow}, {Oliver}, {Owers}, {Pai},
  {Pankratow}, {Parkinson}, {Paschke}, {Paterson}, {Pecontal}, {Parry},
  {Phillips}, {Pillepich}, {Pinard}, {Pirard}, {Piskunov}, {Plank},
  {Pl{\"u}schke}, {Pons}, {Popesso}, {Power}, {Pragt}, {Pramskiy}, {Pryer},
  {Quattri}, {Queiroz}, {Quirrenbach}, {Rahurkar}, {Raichoor}, {Ramstedt},
  {Rau}, {Recio-Blanco}, {Reiss}, {Renaud}, {Revaz}, {Rhode}, {Richard},
  {Richter}, {Rix}, {Robotham}, {Roelfsema}, {Romaniello}, {Rosario},
  {Rothmaier}, {Roukema}, {Ruchti}, {Rupprecht}, {Rybizki}, {Ryde}, {Saar},
  {Sadler}, {Sahl{\'e}n}, {Salvato}, {Sassolas}, {Saunders}, {Saviauk},
  {Sbordone}, {Schmidt}, {Schnurr}, {Scholz}, {Schwope}, {Seifert}, {Shanks},
  {Sheinis}, {Sivov}, {Sk{\'u}lad{\'o}ttir}, {Smartt}, {Smedley}, {Smith},
  {Smith}, {Sorce}, {Spitler}, {Starkenburg}, {Steinmetz}, {Stilz}, {Storm},
  {Sullivan}, {Sutherland}, {Swann}, {Tamone}, {Taylor}, {Teillon}, {Tempel},
  {ter Horst}, {Thi}, {Tolstoy}, {Trager}, {Traven}, {Tremblay}, {Tresse},
  {Valentini}, {van de Weygaert}, {van den Ancker}, {Veljanoski}, {Venkatesan},
  {Wagner}, {Wagner}, {Walcher}, {Waller}, {Walton}, {Wang}, {Winkler},
  {Wisotzki}, {Worley}, {Worseck}, {Xiang}, {Xu}, {Yong}, {Zhao}, {Zheng},
  {Zscheyge}, \& {Zucker}}]{2019Msngr.175....3D}
{de Jong}, R.~S., {Agertz}, O., {Berbel}, A.~A., {et~al.} 2019, The Messenger,
  175, 3, \eprint{1903.02464}

\bibitem[{{DESI Collaboration} {et~al.}(2016){DESI Collaboration}, {Aghamousa},
  {Aguilar}, {Ahlen}, {Alam}, {Allen}, {Allende Prieto}, {Annis}, {Bailey},
  {Balland}, {Ballester}, {Baltay}, {Beaufore}, {Bebek}, {Beers}, {Bell},
  {Bernal}, {Besuner}, {Beutler}, {Blake}, {Bleuler}, {Blomqvist}, {Blum},
  {Bolton}, {Briceno}, {Brooks}, {Brownstein}, {Buckley-Geer}, {Burden},
  {Burtin}, {Busca}, {Cahn}, {Cai}, {Cardiel-Sas}, {Carlberg}, {Carton},
  {Casas}, {Castand er}, {Cervantes-Cota}, {Claybaugh}, {Close}, {Coker},
  {Cole}, {Comparat}, {Cooper}, {Cousinou}, {Crocce}, {Cuby}, {Cunningham},
  {Davis}, {Dawson}, {de la Macorra}, {De Vicente}, {Delubac}, {Derwent},
  {Dey}, {Dhungana}, {Ding}, {Doel}, {Duan}, {Ealet}, {Edelstein},
  {Eftekharzadeh}, {Eisenstein}, {Elliott}, {Escoffier}, {Evatt}, {Fagrelius},
  {Fan}, {Fanning}, {Farahi}, {Farihi}, {Favole}, {Feng}, {Fernandez},
  {Findlay}, {Finkbeiner}, {Fitzpatrick}, {Flaugher}, {Flender}, {Font-Ribera},
  {Forero-Romero}, {Fosalba}, {Frenk}, {Fumagalli}, {Gaensicke}, {Gallo},
  {Garcia-Bellido}, {Gaztanaga}, {Pietro Gentile Fusillo}, {Gerard},
  {Gershkovich}, {Giannantonio}, {Gillet}, {Gonzalez-de-Rivera},
  {Gonzalez-Perez}, {Gott}, {Graur}, {Gutierrez}, {Guy}, {Habib}, {Heetderks},
  {Heetderks}, {Heitmann}, {Hellwing}, {Herrera}, {Ho}, {Holland}, {Honscheid},
  {Huff}, {Hutchinson}, {Huterer}, {Hwang}, {Illa Laguna}, {Ishikawa},
  {Jacobs}, {Jeffrey}, {Jelinsky}, {Jennings}, {Jiang}, {Jimenez}, {Johnson},
  {Joyce}, {Jullo}, {Juneau}, {Kama}, {Karcher}, {Karkar}, {Kehoe}, {Kennamer},
  {Kent}, {Kilbinger}, {Kim}, {Kirkby}, {Kisner}, {Kitanidis}, {Kneib},
  {Koposov}, {Kovacs}, {Koyama}, {Kremin}, {Kron}, {Kronig}, {Kueter-Young},
  {Lacey}, {Lafever}, {Lahav}, {Lambert}, {Lampton}, {Land riau}, {Lang},
  {Lauer}, {Le Goff}, {Le Guillou}, {Le Van Suu}, {Lee}, {Lee}, {Leitner},
  {Lesser}, {Levi}, {L'Huillier}, {Li}, {Liang}, {Lin}, {Linder}, {Loebman},
  {Luki{\'c}}, {Ma}, {MacCrann}, {Magneville}, {Makarem}, {Manera}, {Manser},
  {Marshall}, {Martini}, {Massey}, {Matheson}, {McCauley}, {McDonald},
  {McGreer}, {Meisner}, {Metcalfe}, {Miller}, {Miquel}, {Moustakas}, {Myers},
  {Naik}, {Newman}, {Nichol}, {Nicola}, {Nicolati da Costa}, {Nie}, {Niz},
  {Norberg}, {Nord}, {Norman}, {Nugent}, {O'Brien}, {Oh}, {Olsen}, {Padilla},
  {Padmanabhan}, {Padmanabhan}, {Palanque-Delabrouille}, {Palmese},
  {Pappalardo}, {P{\^a}ris}, {Park}, {Patej}, {Peacock}, {Peiris}, {Peng},
  {Percival}, {Perruchot}, {Pieri}, {Pogge}, {Pollack}, {Poppett}, {Prada},
  {Prakash}, {Probst}, {Rabinowitz}, {Raichoor}, {Ree}, {Refregier}, {Regal},
  {Reid}, {Reil}, {Rezaie}, {Rockosi}, {Roe}, {Ronayette}, {Roodman}, {Ross},
  {Ross}, {Rossi}, {Rozo}, {Ruhlmann-Kleider}, {Rykoff}, {Sabiu}, {Samushia},
  {Sanchez}, {Sanchez}, {Schlegel}, {Schneider}, {Schubnell}, {Secroun},
  {Seljak}, {Seo}, {Serrano}, {Shafieloo}, {Shan}, {Sharples}, {Sholl},
  {Shourt}, {Silber}, {Silva}, {Sirk}, {Slosar}, {Smith}, {Smoot}, {Som},
  {Song}, {Sprayberry}, {Staten}, {Stefanik}, {Tarle}, {Sien Tie}, {Tinker},
  {Tojeiro}, {Valdes}, {Valenzuela}, {Valluri}, {Vargas-Magana}, {Verde},
  {Walker}, {Wang}, {Wang}, {Weaver}, {Weaverdyck}, {Wechsler}, {Weinberg},
  {White}, {Yang}, {Yeche}, {Zhang}, {Zhao}, {Zheng}, {Zhou}, {Zhou}, {Zhu},
  {Zou}, \& {Zu}}]{2016arXiv161100036D}
{DESI Collaboration}, {Aghamousa}, A., {Aguilar}, J., {et~al.} 2016, arXiv
  e-prints, arXiv:1611.00036, \eprint{1611.00036}

\bibitem[{{Driver} {et~al.}(2019){Driver}, {Liske}, {Davies}, {Robotham},
  {Baldry}, {Brown}, {Cluver}, {Kuijken}, {Loveday}, {McMahon}, {Meyer},
  {Norberg}, {Owers}, {Power}, {Taylor}, \& {WAVES Team}}]{2019Msngr.175...46D}
{Driver}, S.~P., {Liske}, J., {Davies}, L.~J.~M., {et~al.} 2019, The Messenger,
  175, 46, \eprint{1903.02473}

\bibitem[{{Finoguenov} {et~al.}(2019){Finoguenov}, {Merloni}, {Comparat},
  {Nandra}, {Salvato}, {Tempel}, {Raichoor}, {Richard}, {Kneib}, {Pillepich},
  {Sahl{\'e}n}, {Popesso}, {Norberg}, {McMahon}, \& {4MOST
  Collaboration}}]{2019Msngr.175...39F}
{Finoguenov}, A., {Merloni}, A., {Comparat}, J., {et~al.} 2019, The Messenger,
  175, 39, \eprint{1903.02471}

\bibitem[{{Guiglion} {et~al.}(2019){Guiglion}, {Battistini}, {Bell}, {Bensby},
  {Boller}, {Chiappini}, {Comparat}, {Christlieb}, {Church}, {Cioni}, {Davies},
  {Dwelly}, {de Jong}, {Feltzing}, {Gueguen}, {Howes}, {Irwin}, {Kushniruk},
  {Lam}, {Liske}, {McMahon}, {Merloni}, {Norberg}, {Robotham}, {Schnurr},
  {Sorce}, {Starkenburg}, {Storm}, {Swann}, {Tempel}, {Thi}, {Worley},
  {Walcher}, \& {4MOST Collaboration}}]{2019Msngr.175...17G}
{Guiglion}, G., {Battistini}, C., {Bell}, C.~P.~M., {et~al.} 2019, The
  Messenger, 175, 17, \eprint{1903.02466}

\bibitem[{{Hansen} {et~al.}(2015){Hansen}, {Ludwig}, {Seifert}, {Koch}, {Xu},
  {Caffau}, {Christlieb}, {Korn}, {Lind}, {Sbordone}, {Ruchti}, {Feltzing}, {de
  Jong}, \& {Barden}}]{2015AN....336..665H}
{Hansen}, C.~J., {Ludwig}, H.~G., {Seifert}, W., {et~al.} 2015, Astronomische
  Nachrichten, 336, 665, \eprint{1508.02714}

\bibitem[{{Helmi} {et~al.}(2019){Helmi}, {Irwin}, {Deason}, {Balbinot},
  {Belokurov}, {Bland-Hawthorn}, {Christlieb}, {Cioni}, {Feltzing}, {Grebel},
  {Kordopatis}, {Starkenburg}, {Walton}, \& {Worley}}]{2019Msngr.175...23H}
{Helmi}, A., {Irwin}, M., {Deason}, A., {et~al.} 2019, The Messenger, 175, 23,
  \eprint{1903.02467}

\bibitem[{{Klypin} {et~al.}(2016){Klypin}, {Yepes}, {Gottl{\"o}ber}, {Prada},
  \& {He{\ss}}}]{2016MNRAS.457.4340K}
{Klypin}, A., {Yepes}, G., {Gottl{\"o}ber}, S., {Prada}, F., \& {He{\ss}}, S.
  2016, \mnras, 457, 4340, \eprint{1411.4001}

\bibitem[{{Merloni} {et~al.}(2019){Merloni}, {Alexander}, {Banerji}, {Boller},
  {Comparat}, {Dwelly}, {Fotopoulou}, {McMahon}, {Nandra}, {Salvato}, {Croom},
  {Finoguenov}, {Krumpe}, {Lamer}, {Rosario}, {Schwope}, {Shanks}, {Steinmetz},
  {Wisotzki}, \& {Worseck}}]{2019Msngr.175...42M}
{Merloni}, A., {Alexander}, D.~A., {Banerji}, M., {et~al.} 2019, The Messenger,
  175, 42, \eprint{1903.02472}

\bibitem[{{Richard} {et~al.}(2019){Richard}, {Kneib}, {Blake}, {Raichoor},
  {Comparat}, {Shanks}, {Sorce}, {Sahl{\'e}n}, {Howlett}, {Tempel}, {McMahon},
  {Bilicki}, {Roukema}, {Loveday}, {Pryer}, {Buchert}, {Zhao}, \& {CRS
  Team}}]{2019Msngr.175...50R}
{Richard}, J., {Kneib}, J.-P., {Blake}, C., {et~al.} 2019, The Messenger, 175,
  50, \eprint{1903.02474}

\bibitem[{{Robotham} {et~al.}(2010){Robotham}, {Driver}, {Norberg}, {Baldry},
  {Bamford}, {Hopkins}, {Liske}, {Loveday}, {Peacock}, {Cameron}, {Croom},
  {Doyle}, {Frenk}, {Hill}, {Jones}, {van Kampen}, {Kelvin}, {Kuijken},
  {Nichol}, {Parkinson}, {Popescu}, {Prescott}, {Sharp}, {Sutherland },
  {Thomas}, \& {Tuffs}}]{2010PASA...27...76R}
{Robotham}, A., {Driver}, S.~P., {Norberg}, P., {et~al.} 2010, \pasa, 27, 76,
  \eprint{0910.5121}

\bibitem[{{Sharma} {et~al.}(2011){Sharma}, {Bland-Hawthorn}, {Johnston}, \&
  {Binney}}]{2011ApJ...730....3S}
{Sharma}, S., {Bland-Hawthorn}, J., {Johnston}, K.~V., \& {Binney}, J. 2011,
  \apj, 730, 3, \eprint{1101.3561}

\bibitem[{{Sheinis} {et~al.}(2014){Sheinis}, {Saunders}, {Gillingham},
  {Farrell}, {Muller}, {Smedley}, {Brzeski}, {Waller}, {Gilbert}, \&
  {Smith}}]{2014SPIE.9151E..1XS}
{Sheinis}, A., {Saunders}, W., {Gillingham}, P., {et~al.} 2014, in Society of
  Photo-Optical Instrumentation Engineers (SPIE) Conference Series, Vol. 9151,
  \procspie, 91511X

\bibitem[{{Smith} {et~al.}(2019){Smith}, {He}, {Cole}, {Stothert}, {Norberg},
  {Baugh}, {Bianchi}, {Wilson}, {Brooks}, {Forero-Romero}, {Moustakas},
  {Percival}, {Tarle}, \& {Wechsler}}]{2019MNRAS.484.1285S}
{Smith}, A., {He}, J.-h., {Cole}, S., {et~al.} 2019, \mnras, 484, 1285,
  \eprint{1809.07355}

\bibitem[{{Swann} {et~al.}(2019){Swann}, {Sullivan}, {Carrick}, {Hoenig},
  {Hook}, {Kotak}, {Maguire}, {McMahon}, {Nichol}, \&
  {Smartt}}]{2019Msngr.175...58S}
{Swann}, E., {Sullivan}, M., {Carrick}, J., {et~al.} 2019, The Messenger, 175,
  58, \eprint{1903.02476}

\bibitem[{{Tamura} {et~al.}(2016){Tamura}, {Takato}, {Shimono}, {Moritani},
  {Yabe}, {Ishizuka}, {Ueda}, {Kamata}, {Aghazarian}, {Arnouts}, {Barban},
  {Barkhouser}, {Borges}, {Braun}, {Carr}, {Chabaud}, {Chang}, {Chen}, {Chiba},
  {Chou}, {Chu}, {Cohen}, {de Almeida}, {de Oliveira}, {de Oliveira}, {Dekany},
  {Dohlen}, {dos Santos}, {dos Santos}, {Ellis}, {Fabricius}, {Ferrand},
  {Ferreira}, {Golebiowski}, {Greene}, {Gross}, {Gunn}, {Hammond}, {Harding},
  {Hart}, {Heckman}, {Hirata}, {Ho}, {Hope}, {Hovland}, {Hsu}, {Hu}, {Huang},
  {Jaquet}, {Jing}, {Karr}, {Kimura}, {King}, {Komatsu}, {Le Brun}, {Le
  F{\`e}vre}, {Le Fur}, {Le Mignant}, {Ling}, {Loomis}, {Lupton}, {Madec},
  {Mao}, {Marrara}, {Mendes de Oliveira}, {Minowa}, {Morantz}, {Murayama},
  {Murray}, {Ohyama}, {Orndorff}, {Pascal}, {Pereira}, {Reiley}, {Reinecke},
  {Ritter}, {Roberts}, {Schwochert}, {Seiffert}, {Smee}, {Sodre}, {Spergel},
  {Steinkraus}, {Strauss}, {Surace}, {Suto}, {Suzuki}, {Swinbank}, {Tait},
  {Takada}, {Tamura}, {Tanaka}, {Tresse}, {Verducci}, {Vibert}, {Vidal},
  {Wang}, {Wen}, {Yan}, \& {Yasuda}}]{2016SPIE.9908E..1MT}
{Tamura}, N., {Takato}, N., {Shimono}, A., {et~al.} 2016, in Society of
  Photo-Optical Instrumentation Engineers (SPIE) Conference Series, Vol. 9908,
  Ground-based and Airborne Instrumentation for Astronomy VI, 99081M,
  \eprint{1608.01075}

\bibitem[{{The MSE Science Team} {et~al.}(2019){The MSE Science Team},
  {Babusiaux}, {Bergemann}, {Burgasser}, {Ellison}, {Haggard}, {Huber},
  {Kaplinghat}, {Li}, {Marshall}, {Martell}, {McConnachie}, {Percival},
  {Robotham}, {Shen}, {Thirupathi}, {Tran}, {Yeche}, {Yong}, {Adibekyan},
  {Silva Aguirre}, {Angelou}, {Asplund}, {Balogh}, {Banerjee}, {Bannister},
  {Barr{\'\i}a}, {Battaglia}, {Bayo}, {Bechtol}, {Beck}, {Beers}, {Bellinger},
  {Berg}, {Bestenlehner}, {Bilicki}, {Bitsch}, {Bland-Hawthorn}, {Bolton},
  {Boselli}, {Bovy}, {Bragaglia}, {Buzasi}, {Caffau}, {Cami}, {Carleton},
  {Casagrande}, {Cassisi}, {Catelan}, {Chang}, {Cortese}, {Damjanov}, {Davies},
  {de Grijs}, {de Rosa}, {Deason}, {di Matteo}, {Drlica-Wagner}, {Erkal},
  {Escorza}, {Ferrarese}, {Fleming}, {Font-Ribera}, {Freeman}, {G{\"a}nsicke},
  {Gabdeev}, {Gallagher}, {Gandolfi}, {Garc{\'\i}a}, {Gaulme}, {Geha},
  {Gennaro}, {Gieles}, {Gilbert}, {Gordon}, {Goswami}, {Greco}, {Grillmair},
  {Guiglion}, {H{\'e}nault-Brunet}, {Hall}, {Hand ler}, {Hansen}, {Hathi},
  {Hatzidimitriou}, {Haywood}, {Hern{\'a}ndez Santisteban}, {Hillenbrand},
  {Hopkins}, {Howlett}, {Hudson}, {Ibata}, {Ili{\'c}}, {Jablonka}, {Ji},
  {Jiang}, {Juneau}, {Karakas}, {Karinkuzhi}, {Kim}, {Kong}, {Konstantopoulos},
  {Krogager}, {Lagos}, {Lallement}, {Laporte}, {Lebreton}, {Lee}, {Lewis},
  {Lianou}, {Liu}, {Lodieu}, {Loveday}, {M{\'e}sz{\'a}ros}, {Makler}, {Mao},
  {Marchesini}, {Martin}, {Mateo}, {Melis}, {Merle}, {Miglio}, {Gohar
  Mohammad}, {Molaverdikhani}, {Monier}, {Morel}, {Mosser}, {Nataf}, {Necib},
  {Neilson}, {Newman}, {Nierenberg}, {Nord}, {Noterdaeme}, {O'Dea}, {Oshagh},
  {Pace}, {Palanque-Delabrouille}, {Pandey}, {Parker}, {Pawlowski}, {Peter},
  {Petitjean}, {Petric}, {Placco}, {Popovi{\'c}}, {Price-Whelan}, {Prsa},
  {Ravindranath}, {Rich}, {Ruan}, {Rybizki}, {Sakari}, {Sanderson}, {Schiavon},
  {Schimd}, {Serenelli}, {Siebert}, {Siudek}, {Smiljanic}, {Smith}, {Sobeck},
  {Starkenburg}, {Stello}, {Szab{\'o}}, {Szabo}, {Taylor}, {Thanjavur},
  {Thomas}, {Tollerud}, {Toonen}, {Tremblay}, {Tresse}, {Tsantaki},
  {Valentini}, {Van Eck}, {Variu}, {Venn}, {Villaver}, {Walker}, {Wang},
  {Wang}, {Wilson}, {Wright}, {Xu}, {Yildiz}, {Zhang}, {Zwintz}, {Anguiano},
  {Bedell}, {Chaplin}, {Collet}, {Cuillandre}, {Duc}, {Flagey}, {Hermes},
  {Hill}, {Kamath}, {Laychak}, {Ma{\l}ek}, {Marley}, {Sheinis}, {Simons},
  {Sousa}, {Szeto}, {Ting}, {Vegetti}, {Wells}, {Babas}, {Bauman}, {Bosselli},
  {C{\^o}t{\'e}}, {Colless}, {Comparat}, {Courtois}, {Crampton}, {Croom},
  {Davies}, {de Grijs}, {Denny}, {Devost}, {di Matteo}, {Driver},
  {Fernandez-Lorenzo}, {Guhathakurta}, {Han}, {Higgs}, {Hill}, {Ho}, {Hopkins},
  {Hudson}, {Ibata}, {Isani}, {Jarvis}, {Johnson}, {Jullo}, {Kaiser}, {Kneib},
  {Koda}, {Koshy}, {Mignot}, {Murowinski}, {Newman}, {Nusser}, {Pancoast},
  {Peng}, {Peroux}, {Pichon}, {Poggianti}, {Richard}, {Salmon}, {Seibert},
  {Shastri}, {Smith}, {Sutaria}, {Tao}, {Taylor}, {Tully}, {van Waerbeke},
  {Vermeulen}, {Walker}, {Willis}, {Willot}, \&
  {Withington}}]{2019arXiv190404907T}
{The MSE Science Team}, {Babusiaux}, C., {Bergemann}, M., {et~al.} 2019, arXiv
  e-prints, arXiv:1904.04907, \eprint{1904.04907}

\bibitem[{{Walcher} {et~al.}(2019){Walcher}, {Banerji}, {Battistini}, {Bell},
  {Bellido-Tirado}, {Bensby}, {Bestenlehner}, {Boller}, {Brynnel}, {Casey},
  {Chiappini}, {Christlieb}, {Church}, {Cioni}, {Croom}, {Comparat}, {Davies},
  {de Jong}, {Dwelly}, {Enke}, {Feltzing}, {Feuillet}, {Fouesneau}, {Ford},
  {Frey}, {Gonzalez-Solares}, {Gueguen}, {Howes}, {Irwin}, {Klar},
  {Kordopatis}, {Korn}, {Krumpe}, {Kushniruk}, {Lam}, {Lewis}, {Lind}, {Liske},
  {Loveday}, {Mainieri}, {Martell}, {Matijevic}, {McMahon}, {Merloni},
  {Murphy}, {Niederhofer}, {Norberg}, {Pramskiy}, {Romaniello}, {Robotham},
  {Rothmaier}, {Ruchti}, {Schnurr}, {Schwope}, {Smedley}, {Sorce},
  {Starkenburg}, {Stilz}, {Storm}, {Tempel}, {Thi}, {Traven}, {Valentini}, {van
  den Ancker}, {Walton}, {Winkler}, \& {Worley}}]{2019Msngr.175...12W}
{Walcher}, C.~J., {Banerji}, M., {Battistini}, C., {et~al.} 2019, The
  Messenger, 175, 12, \eprint{1903.02465}

\end{thebibliography}

\end{document}